\documentclass[sigconf]{acmart}

\AtBeginDocument{%
  \providecommand\BibTeX{{%
    \normalfont B\kern-0.5em{\scshape i\kern-0.25em b}\kern-0.8em\TeX}}}
    
\newcommand{\sdhfix}[1]{{\color{red}[sdh: #1]}}
\newcommand{\bbfix}[1]{{\color{blue}[bb: #1]}}
\usepackage{tikz}
\usepackage{booktabs}
\usepackage{tabularx}

\copyrightyear{2022}
\acmYear{2022}
\setcopyright{rightsretained}
\acmConference[AIES'22]{Proceedings of the 2022 AAAI/ACM Conference on AI, Ethics, and Society}{August 1--3, 2022}{Oxford, United Kingdom}
\acmBooktitle{Proceedings of the 2022 AAAI/ACM Conference on AI, Ethics, and Society (AIES'22), August 1--3, 2022, Oxford, United Kingdom}\acmDOI{10.1145/3514094.3534146}
\acmISBN{978-1-4503-9247-1/22/08}
\begin{document}
\fancyhead{}

\title{Current and Near-Term AI as a Potential Existential Risk Factor}

\author{Benjamin S. Bucknall}
\authornote{Corresponding Author. This work was conducted while BSB was a research intern at the Existential Risk Observatory, Amsterdam.}
\affiliation{%
  \institution{Department of Information Technology \\ Uppsala University}
  \country{Sweden}
}
\email{ben.s.bucknall@gmail.com}

\author{Shiri Dori-Hacohen}
\authornote{Senior Corresponding Author}
\affiliation{%
  \institution{Reducing Information Ecosystem Threats (RIET) Lab \\ Computer Science \& Engineering Department \\ University of Connecticut}
  \country{USA}
}
\email{shiridh@uconn.edu}

%\renewcommand{\shortauthors}{Bucknall and Dori-Hacohen}

\begin{abstract}
There is a substantial and ever-growing corpus of evidence and literature exploring the impacts of Artificial intelligence (AI) technologies on society, politics, and humanity as a whole. A separate, parallel body of work has explored existential risks to humanity, including but not limited to that stemming from unaligned Artificial General Intelligence (AGI). In this paper, we problematise the notion that current and near-term artificial intelligence technologies have the potential to contribute to existential risk by acting as intermediate risk factors, and that this potential is not limited to the unaligned AGI scenario. We propose the hypothesis that certain already-documented effects of AI can act as existential risk factors, magnifying the likelihood of previously identified sources of existential risk. Moreover, future developments in the coming decade hold the potential to significantly exacerbate these risk factors, even in the absence of artificial general intelligence. Our main contribution is a (non-exhaustive) exposition of potential AI risk factors and the causal relationships between them, focusing on how AI can affect power dynamics and information security. This exposition demonstrates that there exist causal pathways from AI systems to existential risks that do not presuppose hypothetical future AI capabilities. 
\end{abstract}




\maketitle

\section{Introduction}
%In this paper we aim to illuminate and draw attention to how current Artificial Intelligence (AI) technologies can contribute to existential risk through their acting as an existential risk factor. It is clear that AI is having a drastic affect on current global politics, and there is a large and ever-growing corpus of existing literature exploring these effects. However, the central hypothesis of the present paper is that some of these observed effects can act as existential risk factors, magnifying the likelihood of existing sources of existential risk including nuclear-armed conflict, climate change, or indeed the later development of misaligned ASI.

%Importance of existential risks. AI is seen as a big factor, but mainly ASI

As early as the 1950's, leading academics of the time, including Albert Einstein and Bertrand Russell, were warning of risks of human extinction due to the use of nuclear weapons \cite{RussellEinstein}. More recently, the study of risks that may threaten the very existence of our species has grown as an academic discipline following %starting with the publication in 2002 of 
Nick Bostrom's introduction of the topic of existential risks \cite{BostromXRisk}. 
%Since then, %a great deal of effort has been put into analysing the 
Various processes have since been proposed through which our species could go extinct, along with potential approaches to reduce this risk. Crucially, one does not need to estimate any of the existential risk scenarios as highly probable in order to determine that working to prevent them is an extremely valuable prospect; an extremely large negative value with a very low probability, still leads to an incredibly high expected value for even the smallest degree of reduction in risk \cite{BostromXRisk}. %%% TODO: Add citations??
Within the existential risk research community, one of the most discussed risks is that of misaligned artificial intelligence (AI), of which many proposed scenarios rely on the assumption of at least `human-level' artificial general intelligence (AGI), if not outright superintelligence. While we do not deny that some such risks are valid and deserve attention, we feel that less powerful AI systems, including those that are present at the time of writing, ought to also be included in the discussion of existential risks.

%Build on existing digital humanities research, and apply to x-risks
In parallel, much concern and attention is being paid to the shorter-term harms of contemporary AI systems, including in digital humanities~\cite[see, e.g.,][]{chun2021discriminating,crawford2021atlas} and in computer science by the Fairness, Accountability, Transparency and Ethics (FATE) community~\cite[see, e.g.,][]{calders2021introduction}. We find these contributions to be long overdue and highly merited. At the same time, these conversations and studies rarely, if ever, discuss how contemporary AI systems could pose existential threats. This position paper aims to highlight a possible extension to current digital humanities research into the domain of existential risk research, and bridge the gap between the two fields by suggesting plausible pathways through which current and near-term AI systems could impact the existential risk stemming from previously identified sources.

It is important to note that, despite much of the following discussion primarily addressing how AI could increase existential risk, there is also potential for AI to act as a significant positive factor in how we are able to address existential risks. If utilised responsibly, AI is an incredibly powerful tool which would doubtless have many potential applications to devising, coordinating, and implementing responses to the dangers that we face. However, identifying both positive and negative impacts of AI with regards to existential risk, and assessing the net affect of these impacts are not the aims of the present paper, and so discussion of the benefits of AI will be minimal.

%AI could be positive, but we don't talk about that here

This paper will take the following structure. We will begin by briefly summarising relevant prior work on existential risk as well as introducing a number of key pre-existing concepts in the AI and existential risk literatures to which we will refer throughout the paper. The following two sections will each discuss a broad trend of AI, those of AI's impacts on power dynamics and information security respectively, going into detail as to how these trends are currently, and could potentially manifest. We then discuss how each of the commonly discussed sources of existential risk is impacted by the trends discussed in the preceding sections. Following this, a graphical representation of the interacting AI impacts, sources of risks, and risk factors is given and discussed before the final section concludes.

\section{Prior work on Existential Risk}

\begin{comment}

\bbfix{
\begin{itemize}
    \item Cover main ideas of existential risk. Bostrom - Ord
    \item Discuss why x-risk research is important/ethics etc.
    \item Move on to AI as an existential risk, but historically focussed on systems with advanced capabilities. Mention alignment problem
    \item Recently this has expanded to include other types of AI (cite Drexler, TAI), other risks AI could cause (cite Herbert Lin, Christiano), other frames through which to look at the risks (cite Dafoe), and potential solutions (GovAI stuff)
    \item Also, more consideration of "Boring catastrophes"
\end{itemize}}

\sdhfix{Another paper I think it would be really great to cite here is this one I stumbled on \cite{liu2018governing}. here's some text you might want to use, feel free to tweak.
%\end{comment}

In their in-depth exploration of hazards, vulnerabilities and exposures to existential risk, Liu et al.~emphasise the need for expanding our understanding of the variety of paths that might lead to existential risk~\cite{liu2018governing}. Among their arguments, they provide examples of how narrow AI could be sufficient to pose existential risk. In this work, we take our cue from Liu and colleagues, and focus on the potential pathways from narrow AI to existential risk, yet without assuming AGI.
}
\end{comment}

We begin by briefly reviewing the relevant background of existential risk research with a focus on existential risk from AI. As mentioned in the introduction, the study of existential risks to humanity as an academic discipline is said to have begun with Bostrom's influential article \textit{`Existential Risks: Analyzing Human Extinction Scenarios and Related Hazards'} in 2002 \cite{BostromXRisk}, though some discussion was taking place before this \cite{RussellEinstein, Leslie96End}. In his article, Bostrom offers a definition of existential risk to be \textit{`[o]ne where an adverse outcome would either annihilate Earth-originating intelligent life or permanently and drastically curtail its potential.'} He also uses a two-dimensional classification of risks based on both \textit{who} is affected by a risk as well as \textit{how severe} the impact is on each individual that is affected. In terms of these two dimensions, Bostrom takes existential risks to be those that are both \textit{pan-generational}, in that they affect all or almost all living people as well as future generations, and \textit{crushing}, which is the greatest severity that Bostrom considers. Following Bostrom, attention paid to existential risk increased noticeably with several books being published on the subject including Martin Rees' \textit{`Our Final Century?'} \cite{Rees03OurFinalHour}, Richard Posner's \textit{`Catastrophe'} \cite{Posner04Catastrophe}, and Nick Bostrom and Milan \'Cirkovi\'c's \textit{`Global Catastrophic Risks'} \cite{Bostrom08GCRs}.

More recently, in his 2020 book \textit{`The Precipice: Existential Risk and the Future of Humanity'}, Toby Ord summarises much of the work on existential risks carried out since Bostrom's seminal paper \cite{Precipice}. In it he specifies a classification of identified risk sources consisting of: natural risks, such as those from asteroids or supervolcanoes; risks from use of nuclear weapons; biological risks, including pandemics and biotechnology; environmental risks, including those from runaway climate change and biodiversity loss; risks from unaligned artificial intelligence; and unknown future risks. It is Ord's definition of an existential risk as \textit{`a risk that threatens the destruction of humanity's longterm potential'} \cite{Precipice} that we will use in this paper. Note that this definition, in a similar fashion to Bostrom's, includes not only extinction of the human species, but also scenarios of technological or social stagnation, unrecoverable collapses of society or dystopias, all of which would result in the non-fulfilment of our species' potential \cite{LongTermTraj}. Furthermore, we will make use of another definition that Ord highlights, that of an \textit{existential risk factor} (or simply \textit{risk factor}), being a situation, state of affairs, or event that, while may not directly constitute an existential risk, can increase the probability of an existential catastrophe occurring, or reduce our abilities to respond to one \cite{Precipice}. A commonly cited illustrative example is that of a great power war, that is, a war between technologically advanced and powerful nations. While the war itself might not be a source of existential risk, it is certainly plausible that it would contribute to overall risk during, and perhaps after, the war if, for example, it pushes involved nations towards use or development of nuclear, chemical, or biological weapons of mass destruction, or incentivises greater funding towards developing novel, more dangerous weapons technologies.

One of the sources of risk that is estimated to contribute the most to the total amount of risk currently faced by humanity is that of unaligned artificial intelligence, which Ord estimates to pose a one-in-ten chance of existential catastrophe in the coming century. Several books including Bostroms \textit{`Superintelligence'} \cite{Superintelligence}, Stuart Russell's \textit{`Human Compatible'} \cite{HumanCompatible}, and Brian Chrisitian's \textit{`The Alignment Problem'} \cite{Christian21Alignment}, as well as numerous articles \cite[see, e.g.,][]{Amodei16Concrete, Everitt18AGISafety, Hendrycks21Unsolved} have addressed the \textit{alignment problem} of how to ensure that the values of any advanced AI systems developed in the coming years, decades, or centuries are aligned with those of our species.

Much of this literature has started from the assumption of currently hypothetical levels of AI capability, introducing terms such as \textit{superintelligence}, \textit{artificial general intelligence}, or \textit{human-level machine intelligence}, and proceeding to explore the inherent difficulties in controlling such systems and presenting potential strategies for doing so. In recent years however, more focus has been payed to the impacts of AI that are not assumed to have such strong capabilities, and the ways in which such systems could still pose an existential risk. For example, K. Eric Drexler \cite{DrexlerReframing} considers risks from pluralities of interacting AI services, shedding light on a new trajectory to existential catastrophe from AI. Furthermore, researchers such as Paul Christiano \cite{WhatFailureLooksLike} and Joseph Carlsmith \cite{Carlsmith} are now considering risks from \textit{`power-seeking AI'}, that is, systems that despite not being assumed to be at some specific level of generality or intelligence, nonetheless could seek to increase their level of influence for their own strategic advantage. Finally, scholars within the field of AI governance are considering ways in which even weaker conceptions of future AI can have dramatic impacts \cite[see, e.g.,][]{Gruetzemacher22Transformative, Whittlestone21Societal}. For example, Zwetsloot and Dafoe introduce the concept of \textit{structural} risks, complementing the %more commonplace 
notions of \textit{accident} and \textit{misuse} risks from AI \cite{Zwetsloot19Thinking, DafoeAIGov}. Whereas misuse risks are those caused by the deliberate use of AI in a harmful manner, and accident risks occur due to some glitch, fault, or oversight that causes an AI system to exhibit unexpected harmful behaviour, structural risks of AI are those caused by how a system \textit{`shapes the broader environment in ways that could be disruptive or harmful'.} %In a similar vein, 
Dafoe also defines a set of perspectives through which to view the challenges of governing AI. These are the \textit{ecology} and \textit{general purpose technology (GPT)} perspectives, which consider a potential global ecosystem of interacting AI systems, and view AI as a broad GPT comparable to the combustion engine, or electricity, respectively, as well as the more familiar \textit{superintelligence} perspective, which focuses on the challenges of governing AI with cognitive abilities far greater than our own. Compared to the superintelligence perspective, the ecology and GPT perspectives lend themselves to a broader outlook of AI and thus lend themselves to studying potential structural risks from AI.

This shift towards less dramatic and speculative failure scenarios due to advanced AI is accompanied by a similarly growing body that moves away from characterising other existential risks solely as singular events and towards descriptions of risks as the result of the complex interactions between multiple, more mundane vulnerabilities in our social and political systems \cite[see, e.g.,][]{Avin18Classifying, Cotton-Barratt20Defence, liu2018governing}. For example, in their in-depth exploration of hazards, vulnerabilities and exposures to existential risk, Liu et al.~emphasise the need for expanding our understanding of the variety of paths that might lead to existential risk~\cite{liu2018governing}. Among their arguments, they provide examples of how narrow AI could be sufficient to pose existential risk. This work aims to contribute further to this perspective by focusing on the potential indirect pathways from narrow AI to existential risk, without assuming AGI.

\subsection{A note on terminology}
Throughout this paper we make use of Baum's definitions of near-term, mid-term, and long-term AI \cite{BaumMidTerm}. As defined by Baum, near-term AI is \textit{`AI that already exists or is actively under development with a  clear path to being built and deployed'}, while long-term AI is \textit{`AI that has at least human-level general intelligence.'} Mid-term is then loosely defined as AI that falls in between these two categories, with potential for overlap. While some researchers have criticised such distinctions \cite{Prunkl20Beyond}, we believe that these terms can be of instrumental use if defined precisely. Finally, this paper will take a holistic view of AI which considers systems as being situated in, and inseparable from, their wider physical and digital environment. This may include (but is not limited to) political, corporate, and social structures, and the technological systems (software and hardware) through which users interact with AI.

\section{AI Can Shift or Strengthen Existing Power Dynamics}
In this section we consider the ways that AI, as a general purpose technology, can affect the power dynamics between different pairs of actors from nation states, multinational corporations, and the public.
These actors were selected as particularly prominent in relation to AI based on current trends: large, multinational technology companies develop AI for use in their services to be used by the public; the public clearly has an important relationship with nation states; and states interact with multinational technology companies through, for example, regulation. States are also involved in AI research and development through their military and intelligence agencies, and governmental funding of research.

\subsection{State-State Power Relationships}
\label{sec:s-s}
AI has the potential to disturb the relationship between a pair of nation states. Some examples of such disturbance already exist, with voices in the USA and the West more generally expressing increasing concern towards China and Russia's respective goals regarding AI. For example, the National Security Commission on Artificial Intelligence claimed earlier this year that \textit{`America's technological predominance - the backbone of its economic and military power - is under threat'} and that \textit{`AI is deepening the threat posed by cyber attacks and disinformation campaigns that Russia, China, and others are using to infiltrate our society, steal our data, and interfere in our democracy'} \cite{NSCAI}.

%China and Easternisation
In particular, China has undergone staggering development and economic growth since 1979, with its economy doubling in size on every eight years \cite{ChinaGrowth}. Furthermore, the publication in 2017 of the \textit{`New Generation Artificial Intelligence Development Plan'} (AIDP), China's central document outlining it's targets for future AI development, marks a considerable step towards its desired position as a global technology superpower \cite{AIDP}. This dramatic rise in technological capabilities of China and other Asian nations has been interpreted as a broader movement of `Easternisation', that is, a shift of global power towards the East. For example, it has been claimed that \textit{`[i]n the 19th century, the world was Europeanized. In the 20th century, it was Americanized. Now, in the 21st century, the world is being irreversibly Asianized'} \cite{FutureAsian}.

%Regardless of whether or not this is inherently bad, it can have undesireable consequences
%Regardless of whether or not one thinks of such a process of `Easternisation' as inherently dangerous or problematic, there are ways in which it could act as a risk factor.

%Inherently bad (AINow2019)
If one takes the view that such a shift in global power is a primary cause for concern based on moral, political, or social reasoning, then it is clear that this could constitute a risk factor. For example, the AI Now Institute has observed that \textit{`[t]he urgency of ``beating'' China [in terms of the development of AI] is commonly justified based on the nationalist assumption that the US would imbue its AI technologies... with better values than China would. China's authoritarian government is presumed to promote a more dystopian technological future than Western liberal democracies'} \cite{AINOW2019}. If one ascribes to this (admittedly Euro- and US-centric) 
view, it could be argued that there are possible realisations of such a `dystopian technological future' that would constitute an existential catastrophe, regardless of how likely they are.

%Dangerous responses
Nevertheless, it is possible for this global shift to constitute a risk factor even if `Easternisation' is not viewed as a primary risk. The mere existence of such views could itself be a risk if it contributes to a growing sense of an AI arms race between technological superpowers. The emergence of such a competitive dynamic may go on to inhibit international coordination within and without the field of AI, incentivise against AI safety precautions, or apply pressure for investment in intentionally harmful AI technologies such as lethal autonomous weapons (LAWs); some of these trends are already emerging.
It is conceivable that the coming decades might witness an `AI cold war dynamic', increasing the chances of physical conflict between involved states, and diverting resources away from other pressing issues including existential risks.%\begin{footnote}{Though we use US-China relations as an instructive example, we note that this risk could apply to any group of two or more nation-states.}\end{footnote}

\subsection{State-Corporation Power Relationships}
We now consider the evolving nature of relationships between nation states and multinational corporations with significant interests in AI technology.

The past two-decades have seen a monumental rise of private corporations in the technology sector, including Facebook/Meta, Google/Alphabet, Amazon, Microsoft, Netflix, and Twitter, to the extent that, as an example, in the fiscal year ending April 2021, Alphabet had a revenue of \$183bn, which would make it the 54th wealthiest country in terms of GDP.\begin{footnote}{This is a conservative comparison. If market capitalisation is used rather than revenue, the argument is even stronger.}\end{footnote} In addition to their wealth, these corporations interact with the general public to extents unprecedented for private bodies, with a large proportion of our social, professional, and commercial activities being facilitated by services provided by these companies. Technology giants are increasingly coming under scrutiny for the negative externalities they are imposing on their users, such as disinformation, extreme polarization and mental health risks, to name a few~\cite[see, e.g.,][]{lotz2019amazon}. However, the massive scale and disproportionate influence of these corporations can put nation-states in a difficult position when it comes to interacting with them. For example, a proposal made by the Australian government in February 2021 for a new law that would require platforms such as Facebook and Google to pay for the media content that they distribute led to a high-profile dispute. Eventually an agreement was reached, but not before Facebook restricted the actions of Australian news outlets on the site and Google threatened to withdraw its web search service from the country \cite{AusVGoogle}. 

Another difficulty faced by states results from the global reach of these corporations. This is exemplified in the practice of legal tax-avoidance amongst such multinationals, and the resulting attempts of the OECD to adjust international corporation tax prices to better distribute the taxes of multinationals amongst the nations in which they operate. This will be achieved through a plan that will \textit{`ensure a fairer distribution of profits and taxing rights among countries with respect to the largest [multinational enterprises], %MNEs, 
including digital companies'} as well as \textit{`put a floor on competition over corporate income tax, through the introduction of a global minimum corporate tax rate'} \cite{OECD}. 

Inevitably, many of the goals and decisions of such corporations are profit-driven, and as such can often be misaligned with those of wider society. It is thus important for states to devise and enact sufficient regulatory processes in order to ensure that the well-being of the public is prioritised, the functionality of political processes is maintained, and the power of wealthy multinational corporations is used for the benefit of society as a whole; however, this is easier said than done in a rapidly changing technological environment. %If states fail to adequately collaborate on regulation, we may begin to see cases of powerful multinationals utilising their power to advance their potentially misaligned agendas, with significant fallout in the social and political spheres.

Finally, it is worth noting that, much like the international `AI arms race' discussed above, it is %not in
conceivable that a similar situation could occur between a state and a multinational with significant spending in AI research.\footnote{A comparable arms race could also occur between two or more multinational corporations, and may arguably be occurring already.} As in the case of state-state arms races, this could lead to corner-cutting with regards to AI safety, though in contrast to international conflict, at this date it seems unlikely that an arms race involving a private corporation could directly lead to armed conflict.

\subsection{State-Citizen Relationships}
\label{sec:s-cit}
%AI is facilitating greater knowledge of and control over citizens
Next, we consider the changing dynamic between states and their citizens as a result of adopting AI surveillance systems.\footnote{Other AI systems may change the state-citizen dynamic in other ways; we focus here on surveillance which is especially salient for many at the time of writing.} This can take the form of a number of technologies and purposes, including both automatic facial and voice recognition, smart/predictive policing, or the nascent practice of affect recognition, which aims to automatically `read' an individuals emotions from facial micro-expressions. The rapid increase in the ubiquity of such AI systems has been well documented and represents a major increase in the ability that nation-states have to gain knowledge of, and power over, individuals within their borders. The 2018 annual report by the AI Now Institute at New York University claimed that \textit{`[t]he role of AI in widespread surveillance has expanded immensely in the U.S., China, and many other countries worldwide'} \cite{AINOW2018}. They further backed this up a year later in their 2019 report, claiming that \textit{`... despite growing public concern and regulatory action, the rollout of facial recognition and other risky AI technologies has barely slowed down'} \cite{AINOW2019}. These findings are also supported by Steven Feldstein at the Carnegie Endowment for International Peace who found that \textit{`AI surveillance technology is spreading at a faster rate to a wider range of countries than experts have commonly understood'} \cite{FeldsteinSurveillance}. Feldstein further emphasises that uptake of such systems is not limited to a particular class of state having found example countries in every major world region, and with \textit{`political systems [that] range from closed autocracies to advanced democracies.'} Many of these systems are already highly criticised in the academic community~\cite[see, e.g.,][]{crawford2021atlas,chun2021discriminating} and are in extensive use also by corporations, who have in turn served to normalise this formerly unpalatable\footnote{At least in the West, if not globally.} level of surveillance~\cite{zuboff2019age,crawford2021atlas}. 

%Lack of sufficient legal frameworks, and difficulty of sticking to them
Worryingly, the rise in uptake of AI surveillance technologies is not accompanied by a similarly-paced development in the ethical and legal frameworks that such technologies are embedded in. The AI Now Institute argue there is a growing divide between the theory and practice of ethics in this area~\cite{AINOW2019}. While there has recently been an increase in the awareness of these issues among governments, corporations, and many societal groups, little concrete progress is being made in addressing them. Furthermore, while Feldstein reiterates that some applications of AI surveillance technology are legal%is unlawful
, he also states: \textit{`%The legal standards required to legitimately carry out surveillance are high, and governments struggle to meet them. 
Even democracies with strong rule of law traditions and robust oversight institutions frequently fail to protect individual rights in their surveillance programs'} \cite{FeldsteinSurveillance}. This sets a troubling precedent. If even the most stable and traditionally well-functioning democracies are failing to meet their own regulatory standards for current systems, how will they deal with the more-powerful technologies on the horizon - and furthermore, what hope do we have that less democratic countries will abide by international treaties aiming to enforce ethical use of AI in surveillance?

%Conclusion: Orwellian scenarios are becoming more and more possible
This rapid development of AI surveillance technology, and the lagging ethical and legal frameworks, almost certainly raises the possibility of a 1984-style, Orwellian repressive autocracy (regardless of the actual probability of such a scenario playing out in practice). Under Ord's definition of existential risk the emergence and stabilisation of such a regime would constitute an existential catastrophe.

%Interesting interaction with the private providers of this technology (AINow 2019)
Finally, it is worth mentioning an interesting connection between the above discussion and the previous subsection addressing state-corporation relations. Many of the hardware and software components of AI surveillance systems are developed by private companies that sell the technology to governments and police departments; privatization also legitimises surveillance that would be considered extra-legal if run by the state~\cite{crawford2021atlas}. The AI Now Institute draws a connection between smart city projects and the consolidation of further power in the hands of corporations%, \textit{`So-called ``smart city'' projects around the world are consolidating power over civic life in the hands of for-profit technology companies, putting them in charge of managing critical resources and information.'}
~\cite{AINOW2019}. A concerning example is the ongoing partnerships between Amazon's \textit{`Ring'} doorbell and more than 2,000 U.S. police departments \cite{Ring2000, crawford2021atlas}. Emails released by VICE News show 
\textit{Ring} employees encouraging police departments to share social media posts advertising \textit{Ring} and its partner app \textit{Neighbor}, as well as advising them on how to best persuade hesitant residents to share footage from their \textit{Ring} doorbells~\cite{AmazonCoaching}. The implications to, and extent of, citizen privacy infringements due to public-private surveillance partnerships are only beginning to be understood. %This set-up is concerning both from the perspective of state-corporation relationships discussed above, and the power that state agencies such as police departments have in surveilling their populations.

\section{AI Affects Information Transfer and Access}
In this section we look at the ways that current AI systems, as situated in their wider environments, can affect individuals' and states' access to information and the effect that this could have on social and political systems.

\subsection{Information Ecosystem Threats}%Recommender systems}
\label{sec:InfEco}
The advertising-based revenue model that underlies most of the internet today is driven and propelled by incredibly advanced AI systems, leading to well-documented and troubling phenomena such as disinformation,\footnote{Disinformation campaigns refer to coordinated attempts to spread false information; misinformation is the unintentional sharing of false information~\cite[see, e.g.,][]{dorihacohen2021restoring,ong2018architects}.} extreme polarization, hate speech, and self-radicalization~\cite{hwang2020subprime,zuboff2019age}. Much has been written about how web search and social media, two of the predominant modes of interaction with online information, both rely on AI for their technological and financial success. Underlying the explosive growth of search engine and social media are state-of-the-art AI approaches, including chiefly search and recommender systems, trained on unprecedented amounts of user data and optimised to maximise revenue.

The connections between the ad-tech, AI-powered giants such as Alphabet (Google) and Meta (Facebook) and the meteoric rise of information campaigns, bots, and weaponised controversy have been well documented~\cite[e.g.,][]{mueller2019mueller,hwang2020subprime}. As the first entry in their \textit{`Ledger of Harms'}, the Center for Humane Technology has listed \textit{`Making Sense of the World: Misinformation, conspiracy theories, and fake news'}~\cite{ledger2020}. `Filter bubbles' refer to the feedback loops formed whereby a user is shown search or social media results that align with their existing preferences, leading to ever-increasing levels of confirmation bias~\cite{pariser2011filter,sunstein2018republic}, which significantly exacerbate the pre-internet phenomenon of echo chambers~\cite{EchoChamber}. During their 2016 information campaigns, for example, the Internet Research Agency (IRA) utilised certain features of the Facebook feed algorithm in order to create large-scale groups of like-minded users which were then gradually primed to be more vulnerable to disinformation~\cite{mueller2019mueller}. Optimization of metrics such as `time spent' and `engagement' leads to the prioritization and artificial amplification of emotionally-charged, evocative content, often focused on negative emotions such as anger~\cite{dorihacohen2021restoring} and habits such as `doomscrolling'~\cite{sharma2022dark}, serving as a form of `reward hacking' for the human mind, whereby the users are dehumanised and converted into mere inputs to the AI's profit machine~\cite{crawford2021atlas}. Such optimizations may additionally lead the AI to shift the preferences of its users~\cite{kalimeris2021preference,Krueger2019}, whether to make them more easily predictable~\cite{Russell} or simply to better serve the company's profit motive~\cite{dorihacohen2021restoring}. The extreme polarization that results contributes significantly to the spread of misinformation~\cite{Vicario2016}, and is actively exploited by states and other players sowing disinformation~\cite{mueller2019mueller}, leading to a vicious cycle~\cite{dorihacohen2021restoring,Bak-Colemane2025764118}. This process has led to a breakdown of intersubjectivity in the US and other countries, arguably driven by social media and its underlying AI systems. 
The plot thickens when these challenges coincide and intersect with other, related issues of hate speech, self-radicalization online, incentives for brevity~\cite{Brevity,TwitterDoubleLimit}, and trolling behaviors~\cite{jhaver2018online}, making nuanced conversations all but impossible~\cite{dorihacohen2021restoring}. Indeed, a recent and prominent paper published in the Proceedings of the National Academy of Sciences has argued that \textit{`seemingly minor algorithmic decisions'} may be reshaping our long-evolved information-foraging and decision-making processes in as of yet undetermined but potentially harmful ways~\cite{Bak-Colemane2025764118}. 

On top of all these, AI-powered synthetic media -- including, but not limited to, deep fakes -- pose an entirely new level of challenge that society has yet to fully reckon with~\cite{chesney2019deep}, simultaneously allowing the creation of incredibly convincing false narratives, while providing cover frequently referred to as the `liar's dividend' - whereby true and damaging evidence can be waved off as deep fakes~\cite{schiff21liar}. Given the extraordinary success of contemporary disinformation campaigns relying chiefly on so-called ``cheap fakes,'' an overactive imagination is not required in order to see the feasibility of wide-scale AI-powered manipulation in the near future. 

Taken together, these are significant and growing threats to the information ecosystem, and by extension, to the collective decision making capacity of humanity. These threats have the potential to act as a mediating risk factor, particularly regarding those sources of existential risk for which collective action of the public can directly affect the severity of outcomes, such as pandemics, nuclear war and climate change, as we will discuss in further detail below.%\begin{footnote}{This is discussed in more detail in the following section.}\end{footnote}. 

Finally, some scholars have argued that, over and above whatever multiplier effect they may have to other existential risks, information ecosystem threats constitute an existential risk of their own right, painting a dire image of a dystopian, 1984-esque world where truth ceases to carry meaning~\cite{HerbertLin}.

\subsection{Cybersecurity and International Cyber Warfare}
%AI is being increasingly applied in international cybersecurity and cyberwarfare
One of the more natural applications of current and near-term AI is in cybersecurity, as reflected by it's dramatic current and projected growth as an industry worth \$1 billion in 2016 to \$34.8 billion in 2025 \cite{AICyberMarket}. However, there are many open questions regarding the large-scale effects of this trend, not least regarding how the current balance between offence and defence in cybersecurity will be affected by the increased use of AI. Though this is still a disputed topic, though there are persuasive reasons to believe that cybersecurity (including its AI developments) leans towards offense~\cite{JohnsonAI,schneier2018click, Garfinkel19Scale, Buchanan22NewFire}.

%How AI could favour offence
There are many ways in which the use of AI in cybersecurity may make it easier to successfully carry out attacks. %For a start, 
Matteo \textit{et al.} note that the use of machine learning methods in the creation of defense systems creates a further vulnerability in the system itself, if attackers are able to influence the training of the system in use \cite{AICyberSword}. They point to studies showing that carefully constructed changes to training data, imperceptible to human overseers, can result in unexpected behaviour of the trained system\begin{footnote}{Matteo \textit{et al.} give the example of how adding 8\% of faulty data during training of an AI system intended to recommend drug dosages can result in over a 75\% change in the doses for 50\% of patients.}\end{footnote}; a recent survey details many techniques to train machine learning to be robust to such challenges~\cite{duddu2018survey}. %; this effectively becomes a new arms race.
Furthermore, Johnson points out that AI can be used by adversaries to design and implement complex and customised cyber-attacks with unprecedented accuracy and efficiency \cite{JohnsonAI}.  

%How AI could favour defence
On the other hand, AI also provides advanced methods for detecting and responding to cyber-attacks. Wirkuttis and Klein observe that the abilities of AI systems to handle large data sets make them prime candidates for the automation of cybersecurity related tasks, network monitoring, and malicious intrusion identification \cite{AIinCyberSec}. %Furthermore, they explain in more detail how Artificial Neural Networks (ANNs) can be used to monitor network traffic and identify potentially malicious intrusions before an attack is launched.

%It is unknown how AI will affect the current offence/defence balance - opportunities for catastrophe
Cybersecurity between nation-states is now being referred to as `cyberwarfare'~\cite[see, e.g.,][]{duddu2018survey} and actively discussed as `the new cold war'~\cite{acton2020cyber,rohith2019cyber}. To summarise, AI is now a major part of cybersecurity, and an arms race has the potential to lead to potential catastrophes and heightened tensions between nations. %While it is still an open question what large scale impacts the use of AI in cybersecurity will have on the current offence/defence balance, it is highly likely that the balance will shift in some way. Whichever direction this shift goes, national security and intelligence agencies will need to adapt to an ever-evolving landscape if they are to stay ahead of their adversaries. The inherent challenge of responding to the opportunities and challenges that AI presents in this field will open the door to potential catastrophes and heightened tensions between nations.

\section{How These Trends Constitute Risk Factors}
Having discussed some of the potential impacts that current and near-term AI can have, and is already having, % on its physical and digital surroundings, 
we now move on to see how these impacts can act as risk factors. We will begin by discussing general factors that can affect total existential risk from any source, before briefly considering each of the commonly studied sources of risk separately.\footnote{These sources are: nuclear weapons, biotechnology and pandemics, climate change, natural risks, and unaligned artificial general intelligence.}

\subsection{General Risk Factors}
\label{sec:GeneralFactors}
%Less informed public affects democracy
Many of the AI trends and impacts discussed above have direct implications on the functioning of social and political structures. For example, the use of recommender systems and incentives for brevity on social media may lead to an increase in political polarisation and a less informed public respectively. In democratic societies this may have a knock-on effect in terms of the political process. An increase in political polarisation in the general public, for example, may be reflected within political institutions, whereby elected representatives are unwilling to compromise due to a rise in political partisanship. This would lead to an increase in difficulty of passing key legislature aimed at addressing pressing problems, including existential risks, thereby acting as a risk factor. Furthermore, a less informed public may be more susceptible to coordinated mis- and dis-information campaigns, such as the Russian state interference with the 2016 U.S. presidential election~\cite{mueller2019mueller}. Such effects act as risk factors by eroding trust in traditional democratic political structures and processes, thus inhibiting their ability to respond to existential risks. Political processes may also be hindered by changes to state-corporation relations that result in corporations having greater political lobbying power, or political division over government regulation of multinational corporations.

%State-state competition
Secondly, increased state-state competition, in the form of an arms race, AI or otherwise, can also seriously impede responses to existential and catastrophic risks by diverting attention and resources towards the addressing the competition dynamic. While not relating to state-state competition, we can see an example in how more immediately pressing issues can divert attention away from longer-term concerns in countries' reactions to the COVID-19 pandemic. Some writers have observed and pointed out the dangers of tunnel vision towards such scenarios at the expense of other, less immediate but no less serious issues such as climate change and nuclear proliferation, among others%. For example, Peter Giger, writing for the World Economic Forum claims that, in this case, \textit{`[w]e run the risk of diverting our attention exclusively toward COVID-19 issues and losing ground on other major strategic, long-term risks. These may include nuclear proliferation, refugee migration, poverty, cyber security, `biodiversity loss and, of course, climate change.'}
~\cite{Giger}. It is reasonable to expect that the emergence of an AI arms race, or other cold war scenario, could have similar impacts on our responses to other dangers.
%Brief recall of state-citizen dystopia
Finally, as mentioned above, the development and deployment of AI-powered surveillance systems increases the probability % possibility, and thus, probability\begin{footnote}{If a scenario is impossible then the probability of it occurring is clearly 0. Thus, if developments in AI make such scenarios possible, then the probability will increase from 0, regardless of by how much.}\end{footnote},
of Orwellian dystopias of global, sustainable, repressive autocracy.

\subsection{Nuclear Weaponry}
\label{sec:Nuclear}
We now consider near-term AI's impacts on specific sources of risk, starting with that from nuclear weaponry. This is most likely the source of risk that would be most affected by heightened state-state tensions. If competitive dynamics resulting from an AI arms race between two nuclear states grow uncontrolled and become military in nature, this would increase the probability of a first-strike nuclear attack. The evolving international cybersecurity landscape could also play a role in making such an event more likely, if a state is better able to protect its intelligence, or gain greater access to its adversary's. 
Use of nuclear weapons has been widely discussed as a possible source of existential risk due to the possibility of `nuclear winter' \cite{Coupe19NuclearWinter}, %This is a scenario in which the high volumes of smoke resulting from fires following a nuclear strike are lifted to the stratosphere, encircling much of the globe, and blocking large amounts of sunlight from reaching the Earth's surface. Due to the altitude that the smoke will reach, it will not be able to be rained out in the usual water cycle, and thus could remain in place for numerous years. Due to the extreme reduction in sunlight reaching the Earth it is predicted that drastic cooling will occur suddenly, leading to many of the traditional food-growing areas being too cold to produce a reliable harvest. 
which has the potential to lead to %This could result in 
widespread famine, and potential human extinction \cite{Precipice}. Further in-depth discussion of AI's impact on nuclear security can be found in the Stockholm International Peace Research Institute's three-volume report \textit{`The Impact of Artificial Intelligence on Strategic Stability and Nuclear Risk'} \cite{SIPRI19Vol1, SIPRI19Vol2, SIPRI20Vol3}, as well as in \cite{Maas2022MilitaryAI} and \cite{Favaro21Distortion}.

\subsection{Pandemics and Biotechnology}
\label{sec:Bio}
The emergence of a deadly infectious virus, be it naturally occurring or engineered, has been predicted to be a major sources of existential risk~\cite{Precipice, Inglesby19GCBRs}. There are a number of ways in which near-term AI trends can act as a risk factor with regards to pandemics. First, as we have seen from the ongoing COVID-19 pandemic, an adequate and coordinated response to such a scenario requires not only political bodies capable of dealing with the threat, but also a public that will play their part to protect themselves and other individuals from the disease. In this regard, it is likely that our ability to collectively respond %how well we can respond 
to a global pandemic depends on public trust in the relevant political systems, in order to maximise the effectiveness of measures such as quarantining, mask-wearing, and mass vaccination. We have seen waves of COVID misinformation and anti-vaccination conspiracy theories throughout the course of the pandemic, with much of the discourse taking place on social media. If we are to have any hope of %able to better 
responding well to another, potentially more dangerous pandemic in the future, it is crucial that mis- and disinformation and conspiracy theories do not disseminate successfully at anywhere near the scale they have for COVID-19.%kept under control, for which social media platforms will have a significant role to play.

Additionally, and specifically to engineered pandemics, the availability of powerful AI systems may make it easier for malevolent actors to get hold of the technologies and techniques required to design and produce dangerous pathogens to cause a pandemic. While this is admittedly speculative, %how this could be done has not yet been fully explored, we have 
recent developments demonstrate %seen how 
AI's powerful role in solving problems in molecular biology, %was used in the solving of the `protein folding problem', an example of how the power of AI can be applied to great effect in the
namely protein folding~\cite{ProteinFold}. Similarly powerful neural architectures, in malicious hands, may cause untold damage \cite{Urbina22DualUse}.

Finally, heightened international political tensions and state-state rivalry could spur some states to develop and potentially deploy advanced bioweaponry. Indeed, despite the foreseeable widespread condemnation that such an act would provoke, frameworks for regulating such practices is severely under-resourced and underfunded with Ord giving the example that the UN's Biological Weapons Convention operates on an annual budget of only \$1.4 million -- less than the average McDonald's restaurant \cite{Precipice}.

\subsection{Climate Change}
\label{sec:Climate}
Similarly to pandemics, the cumulative impact of individuals' actions are able to impact existential climate change risk either positively \cite{Rolnick22Tackling} or negatively. %have an effect on the amount of risk. 
%However, again similarly to the case of 
Also like pandemics, climate change has a history of being clouded in mis- and disinformation, often perpetuated by those considered to be doing the most damage \cite{BP,oreskes2010merchants}. If we are to be able to address the ongoing climate crisis, the public and political spheres must be able to agree on the dangers that we are facing and the strategies that we can employ to alleviate them. In a society where much of the public's information gathering takes place on social media and where AI algorithms %are able to 
actively promote misleading and hazardous information, AI can increase the likelihood of climate catastrophe.%contribute to the  a large responsibility falls on these platforms to tackle the dangerous societal effects of their services.

Additionally, the practice of developing and training AI systems has a drastic first-hand climate impact, due to the large amounts of energy needed to run the hardware system on which the AI is running~\cite{strubell2019energy}, as well as possible secondary effects through, for example, applications to the exploration and extraction of oil and natural gas~\cite{kaack2020artificial}. It has been estimated that the technology industry as a whole contributed between 3\% and 3.6\% of global greenhouse gas emissions in 2020 \cite{AIEmissions}. Furthermore, it has been reported that training a single natural language processing (NLP) AI model produces 300,000 kilograms of carbon dioxide emissions \cite{strubell2019energy}. According to the AI Now Institute's 2019 report, this amounts to 125 New York to Beijing round-trip flights \cite{AINOW2019}. This is an often-overlooked aspect of the growth of AI that must also be addressed if we are to sustain technological development whilst avoiding a climate catastrophe~\cite{kaack2021aligning}.

\subsection{Natural Risks}
Natural extinction risks are those which humans are not responsible for causing or exacerbating, such as asteroid impacts or supervolcano eruptions. Thus we do not foresee AI to have any significant magnifying impact on such risks aside from the general impacts discussed above.\footnote{However, as noted in the introduction, AI may be useful in mitigating such natural, as well as man-made, risks.}

\subsection{Unaligned AGI}
Finally, we note that developments in near-term AI will have impacts on how the field will proceed far into the future. Thus, both successes and failures of current AI systems may affect the architecture of a potential artificial general intelligence, %or 
the society in which it is developed, or our ability to align it with humanity. For example, if an AI arms race dynamic emerges between states or technology companies, there may be incentives for corner-cutting when it comes to implementing safety practices in order to maximise the capabilities of the system under construction. If such practices continue until %to 
the stage at which an AGI is being developed, %under development,
this could have catastrophic outcomes at the time of deployment of the AGI system.

\section{Discussion}
Figure \ref{fig:CausalGraph} shows a diagrammatic representation of the effects of near-term AI, established sources of existential risk, and the potential risk factors identified in this paper that constitute the causal relations between them. %\bbfix{Replace tikz drawing with normal image once finalised}
\begin{figure*}[h]
\centering
\scalebox{0.7}{
\begin{tikzpicture}[risknode/.style={rectangle, draw=black, fill=red!10, very thick, minimum size=10mm},squarednode/.style={rectangle, draw=black, fill=black!5, very thick, minimum size=10mm},effectnode/.style={rectangle, draw=black, fill=blue!5, very thick, minimum size=10mm}]
	
	% COLUMN 1 ====================================================================
	\node[squarednode, align=center] (AI) at (-9, 0) {Current and \\ Near-term AI};

	% COLUMN 2 ====================================================================
	\node[effectnode, align=center] (Cyber) at (-7, 3.25) {Cybersecurity};
	
	\node[effectnode, align=center] (s-s) at (-4, 3) {State-State \\ Relations};
	
	\node[effectnode, align=center] (s-cor) at (-4, 1.5) {State-Corporation \\ Relations};
	
	\node[effectnode, align=center] (s-cit) at (-4, -1.5) {State-Citizen \\ Relations};
	
	\node[effectnode, align=center] (SocialMedia) at (-4, -3) {Social Media \& \\ Recommender Systems};
	
	\foreach \i in {s-s, s-cor, s-cit, SocialMedia}{
	    \draw[] (AI.east) -- (\i.west);
	}
	\draw[] (AI.east) -- (Cyber.south);
	\draw[] (Cyber.east) -- (s-s.west);

	% COLUMN 3 ====================================================================
	\filldraw[black, fill=yellow!25, very thick, align=center] (-0.5, 1.2) rectangle (2.3, -1);
	\draw[black!50, thick] (-0.5, 0.2) -- (2.3, 0.2);
	\node[align=center, text width = 2cm] at (0.9, 0.7) {Diverted Resources}; 
	\node[align=center, text width = 2.7cm] at (0.9, -0.4) {Compromised \\ Political \\ Decision-making};
	
	\draw[thick, cyan] (s-s.east) -- (-0.5, 0.7);
	
    \draw[thick, magenta] (s-cor.east) -- (-0.5, -0.4);
	    
	\draw[thick, blue] (SocialMedia.east) -- (-0.5, -0.4);

	% COLUMN 4 ====================================================================
	\node[risknode, align=center] (AGI) at (5, 4) {Unaligned \\ AGI};
	\foreach \i in {s-s}{
		\draw[thick, cyan] (\i.east) -- (AGI.west);
	}
	
	\node[risknode] (Natural) at (5, 2.5) {Natural};
	
	\node[risknode] (Nuclear) at (5, 1) {Nuclear};
	
	\draw[thick, orange] plot[smooth, tension=0.5] coordinates {(s-s.east) (1.5, 2.25) (Nuclear.west)};
	
	\node[risknode] (Bio) at (5, -0.5) {Biologicial};
	\foreach \i in {SocialMedia}{
		\draw[thick, blue] (\i.east) -- (Bio.west);
	}
	\draw[thick, orange] plot[smooth, tension=0.5] coordinates {(s-s.east) (2, 1.5) (Bio.west)};
	
	\node[risknode] (Environmental) at (5, -2) {Environmental};
	\foreach \i in {SocialMedia}{
		\draw[thick, blue] (\i.east) -- (Environmental.west);
	}
	
	\node[risknode, align=center] (Orwell) at (5, -3.5) {Stable Repressive \\ Regime};
	\foreach \i in {s-cit}{
		\draw[thick, yellow] (\i.east) -- (Orwell.west);
	}
	
	\foreach \i in {Environmental, Bio, Nuclear, AGI, Natural}{
	    \draw[] (2.3, 0.2) -- (\i.west);
	}
	
	%\draw[thick, red] plot[smooth, tension=0.5] coordinates {(AI.east) (-3, 0) (1, 3) (AGI.west)};
	\draw[thick, red] plot[smooth, tension=0.5] coordinates {(AI.east) (-2.5, -0.5) (0, -1.5) (2.5, -1.5) (Bio.west)};
	\draw[thick, red] plot[smooth, tension=0.7] coordinates {(AI.east) (-6.7, -1.2) (SocialMedia.west)};
	\draw[thick, green] plot[smooth, tension=0.5] coordinates {(AI.east) (-2.75, -0.8) (0, -2) (Environmental.west)};

	% COLUMN 5 ====================================================================
	\node[squarednode] (XRisk) at (9, 0) {Existential Risk};
	
	\foreach \i in {Orwell, Environmental, Bio, Nuclear, AGI, Natural}{
		\draw[] (\i.east) -- (XRisk.west);
	}

\end{tikzpicture}
}
\caption{A graphical representation of the causal pathways from current and near-term AI to existential risk identified in this paper. Blue nodes represent effects of current and near-term AI, whereas red nodes represent identified existential risks \cite{Precipice}. The yellow box represents the general risk factors discussed in Section \ref{sec:GeneralFactors}. Coloured edges represent causal connections as given in Table \ref{tab:EdgeKey}.}
\label{fig:CausalGraph}
\end{figure*}
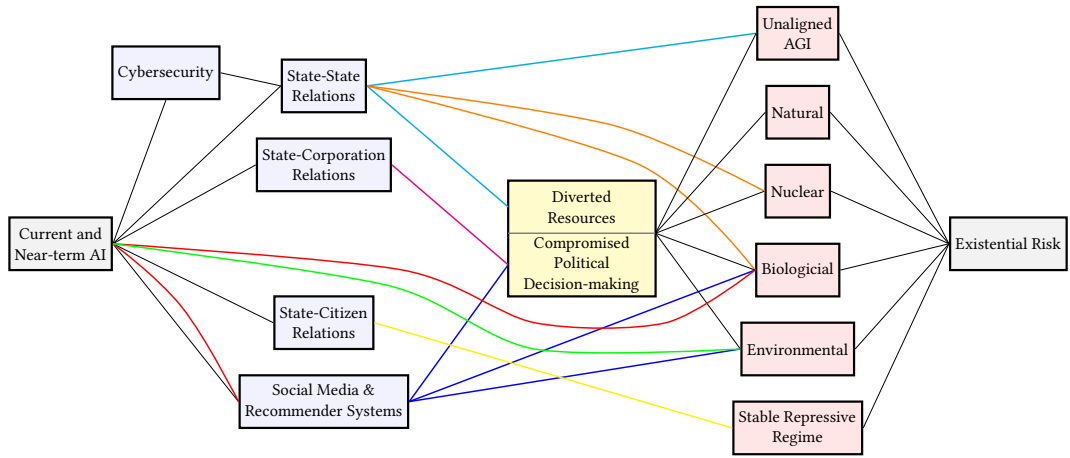
\begin{table}[h]
	\centering
	\caption{Key for edge colours used in Figure \ref{fig:CausalGraph}}
	\label{tab:EdgeKey}
	\begin{tabularx}{\columnwidth}{p{0.1\columnwidth}p{0.45\columnwidth}p{0.35\columnwidth}}
		\toprule
		Edge Colour & Description & References \\
		\midrule
		\textcolor{cyan}{Cyan} & AI arms-race scenario & Section \ref{sec:s-s} \\
		\textcolor{orange}{Orange} & Great power war & Sections \ref{sec:Nuclear} \& \ref{sec:Bio} \\
		\textcolor{red}{Red} & Deliberate malicious use of current AI systems & Section \ref{sec:Bio} \newline \cite{MaliciousAIReport} \\
		\textcolor{green}{Green} & Carbon emissions of training large ML models & Section \ref{sec:Climate} \newline \cite{AIEmissions, AINOW2019, kaack2020artificial, kaack2021aligning, strubell2019energy} \\
		\textcolor{yellow}{Yellow} & AI-enabled surveillance & Section \ref{sec:s-cit} \newline \cite{AINOW2018, AINOW2019, FeldsteinSurveillance, crawford2021atlas, chun2021discriminating, zuboff2019age} \\
		\textcolor{magenta}{Magenta} & Corporate lobbying and government regulation & Section \ref{sec:GeneralFactors} \\
		\textcolor{blue}{Blue} & Modified collective behaviour due to effects on the information ecosystem & Section \ref{sec:InfEco} \newline \cite{Bak-Colemane2025764118, dorihacohen2021restoring, SegerEpistemicSecurity} \\
		\bottomrule
	\end{tabularx}
\end{table}

In this graph we can see explicit causal pathways from current and near-term AI on the left; through its effects on various power relationships, cybersecurity, and the information ecosystem -- represented as blue nodes; to identified sources of existential risk -- given as red nodes. The central yellow box represents general risk factors as given in Section \ref{sec:GeneralFactors}. Edge colours correspond to specific causal relations between nodes, as detailed in Table \ref{tab:EdgeKey}.

We summarise each of these causal relations in turn, starting with cyan edges, representing an AI arms-race scenario (Section \ref{sec:s-s}). Such a scenario could result from heightened state-state tensions and could increase the risk posed by unaligned AGI if the resulting development lacks adequate safety considerations. Furthermore, an arms race could act as a general risk factor through the diversion of resources away from existential risk as states increasingly focus on AI research and development. Accordingly, cyan edges connect state-state relations to the risk from unaligned AGI as well as the general risk factor of diverted resources.

Secondly, orange edges represent a great power war scenario. Such a scenario could also result from heightened geopolitical tensions, and could provide incentive for the development or use of nuclear or biological weapons. Hence, orange edges connect state-state relations to both nuclear and biological risk.

Red edges show the potential malicious use of AI. As noted in Section \ref{sec:Bio}, AI systems could be applied to the engineering of dangerous novel pathogens, and thus a red edge directly connects current and near-term AI to biological risks. Furthermore, AI could be maliciously applied to aid in targeted disinformation campaigns on social media, with the aim of sowing division or confusion in public discourse. Thus, a red edge also connects AI directly to social media. This is alongside a black edge representing effects of AI within social media that are not the result of malicious intent, such as the occurrence of filter bubbles or the spread of misinformation.

There are three causal pathways that only appear as a single edge in Figure \ref{fig:CausalGraph}. Firstly, the green edge represents the direct carbon emissions that result from the training of large ML systems, as discussed in Section \ref{sec:Climate}. This edge thus connects current and near-term AI directly to environmental risks.
Secondly, the yellow edge denotes the use of AI in surveillance technologies, which, as discussed in Section \ref{sec:s-cit}, raises the possibility of stable repressive regimes. Thus it connects the node representing state-citizen relations to the node representing such regimes.
Lastly, the magenta edge shows the practices of corporate lobbying and government regulation of corporations, as mentioned in Section \ref{sec:GeneralFactors}. This edge connects state-corporation relations to the general risk factor of compromised political decision-making.

Finally, the blue edges represent effects on collective behaviour as a result of AI's impact on the information ecosystem. We have discussed how these changes, largely occurring through AI applications in social media, could act as a general risk factor through compromising political decision-making, and can also have effects on the spread of pandemics and the severity of the ongoing climate crisis. Accordingly, blue edges connect social media to general risk factors, as well as biological and environmental risks.

With the complex structure of the interactions between AI effects, risk sources, and risk factors somewhat elucidated by the figure, a few features worthy of discussion are brought to the fore.
Firstly, the position of cybersecurity considerations is fairly unique when compared to the other AI impacts addressed, in that we do not deem it to be a primary risk factor for any of the risks considered. Instead, we consider it to act solely through its impacts on state-state relations through actions noted above such as shifting the offense-defence balance. In this respect, AI's impact on cybersecurity could be considered purely as a constituent component of its impact on state-state relations. While effects on cybersecurity may also impact other parts of the diagram, such as state-citizen relations,\footnote{We thank an anonymous review for this observation.} we do not deem these to be of the scale of constituting an existential risk factor and so are not explicitly included.

Furthermore, Figure \ref{fig:CausalGraph} allows us to crudely compare the relative potential magnitudes that each AI impact contributes to total existential risk. If an AI impact acts as multiple risk factors (corresponding to multiple colours of edges), or acts upon multiple risk sources, this could indicate that this impact affects the magnitude of total existential risk that we face more than those impacts only acting as a single factor, or that only affect a single source. This clearly depends on the magnitudes of each risk factor, as well as the estimated contribution of each risk source to total existential risk, and so any such deductions from this graph are speculative at best. Future work aiming to make this diagram more quantitative by including numerical estimates of risk could be of value, though making such an extension would clearly be both challenging and speculative. That being said, preliminary judgement would deem the effects of AI on state-state relationships to be particularly important for future research as a result of state-state tensions acting on five of the six identified sources of risk through the potential scenarios of an AI arms-race and use of either nuclear or biological weaponry. On the other hand, AI's impact on state-citizen relations seems to be of less concern due to its affecting only the risk of a stable repressive regime through the use of AI in surveillance technologies. We reiterate that such conclusions are highly uncertain and could benefit from quantitative extensions to this framework.

\section{Conclusions and Future Work}
In this paper we have argued that societal and political issues surrounding contemporary AI systems can have far-reaching impacts on humanity through their potential acting as existential risk factors, rather than solely through the development of unaligned AGI. We argued that short-term harms from extant AI systems may magnify, complicate, or exacerbate other existential risks, over and above the harms they are inflicting on present society. In this manner, we have offered a bridge connecting two seemingly distinct areas of study: AI's present harms to society and AI-driven existential risk. By proposing concrete mechanisms for current and near-term AI to act as an intermediate factor to existential risk, we have taken the first step in demonstrating the connection between its primary impacts and existential risk. We believe that more research with the purpose of shedding light on these connections would be incredibly valuable. In particular, we see opportunities to extend the framework presented in Figure \ref{fig:CausalGraph} either through quantitative estimates of relative likelihoods, or qualitative extensions to other AI impacts or sources of existential risk. The current position paper serves only as a first step in identifying and addressing risks of this nature.

\begin{acks}
The authors would like to thank Jos\'e Hern\'andez-Orallo and Gabriel Pedroza for valuable comments and discussion, and Otto Barten for his continued support via the Existential Risk Observatory. We also thank our anonymous reviewers at SafeAI workshop and AIES for detailed feedback, comments, and further reading suggestions. Finally, we thank AI Safety Support for facilitating our initial connection.
\end{acks}

\bibliographystyle{ACM-Reference-Format}
\bibliography{mybib}


\begin{thebibliography}{91}

%%% the Web address.

\ifx \showCODEN    \undefined \def \showCODEN     #1{\unskip}     \fi
\ifx \showDOI      \undefined \def \showDOI       #1{#1}\fi
\ifx \showISBNx    \undefined \def \showISBNx     #1{\unskip}     \fi
\ifx \showISBNxiii \undefined \def \showISBNxiii  #1{\unskip}     \fi
\ifx \showISSN     \undefined \def \showISSN      #1{\unskip}     \fi
\ifx \showLCCN     \undefined \def \showLCCN      #1{\unskip}     \fi
\ifx \shownote     \undefined \def \shownote      #1{#1}          \fi
\ifx \showarticletitle \undefined \def \showarticletitle #1{#1}   \fi
\ifx \showURL      \undefined \def \showURL       {\relax}        \fi
% invisible to TeX
\providecommand\bibfield[2]{#2}
\providecommand\bibinfo[2]{#2}
\providecommand\natexlab[1]{#1}
\providecommand\showeprint[2][]{arXiv:#2}

\bibitem[Acton(2020)]%
        {acton2020cyber}
\bibfield{author}{\bibinfo{person}{James~M Acton}.}
  \bibinfo{year}{2020}\natexlab{}.
\newblock \showarticletitle{Cyber Warfare \& Inadvertent Escalation}.
\newblock \bibinfo{journal}{\emph{D{\ae}dalus}} \bibinfo{volume}{149},
  \bibinfo{number}{2} (\bibinfo{year}{2020}), \bibinfo{pages}{133--149}.
\newblock


\bibitem[Amodei et~al\mbox{.}(2016)]%
        {Amodei16Concrete}
\bibfield{author}{\bibinfo{person}{Dario Amodei}, \bibinfo{person}{Chris Olah},
  \bibinfo{person}{Jacob Steinhardt}, \bibinfo{person}{Paul Christiano},
  \bibinfo{person}{John Schulman}, {and} \bibinfo{person}{Dan Man\'e}.}
  \bibinfo{year}{2016}\natexlab{}.
\newblock \bibinfo{title}{Concrete Problems in {AI} Safety}.
\newblock
\newblock
\showeprint[arxiv]{1606.06565}~[cs.AI]


\bibitem[Avin et~al\mbox{.}(2018)]%
        {Avin18Classifying}
\bibfield{author}{\bibinfo{person}{Shahar Avin}, \bibinfo{person}{Bonnie~C.
  Wintle}, \bibinfo{person}{Julius Weitzdörfer}, \bibinfo{person}{Seán~S. {Ó
  hÉigeartaigh}}, \bibinfo{person}{William~J. Sutherland}, {and}
  \bibinfo{person}{Martin~J. Rees}.} \bibinfo{year}{2018}\natexlab{}.
\newblock \showarticletitle{Classifying global catastrophic risks}.
\newblock \bibinfo{journal}{\emph{Futures}}  \bibinfo{volume}{102}
  (\bibinfo{year}{2018}), \bibinfo{pages}{20--26}.
\newblock
\showISSN{0016-3287}
\urldef\tempurl%
\url{https://doi.org/10.1016/j.futures.2018.02.001}
\showDOI{\tempurl}


\bibitem[Bak-Coleman et~al\mbox{.}(2021)]%
        {Bak-Colemane2025764118}
\bibfield{author}{\bibinfo{person}{Joseph~B. Bak-Coleman},
  \bibinfo{person}{Mark Alfano}, \bibinfo{person}{Wolfram Barfuss},
  \bibinfo{person}{Carl~T. Bergstrom}, \bibinfo{person}{Miguel~A. Centeno},
  \bibinfo{person}{Iain~D. Couzin}, \bibinfo{person}{Jonathan~F. Donges},
  \bibinfo{person}{Mirta Galesic}, \bibinfo{person}{Andrew~S. Gersick},
  \bibinfo{person}{Jennifer Jacquet}, \bibinfo{person}{Albert~B. Kao},
  \bibinfo{person}{Rachel~E. Moran}, \bibinfo{person}{Pawel Romanczuk},
  \bibinfo{person}{Daniel~I. Rubenstein}, \bibinfo{person}{Kaia~J. Tombak},
  \bibinfo{person}{Jay~J. Van~Bavel}, {and} \bibinfo{person}{Elke~U. Weber}.}
  \bibinfo{year}{2021}\natexlab{}.
\newblock \showarticletitle{Stewardship of global collective behavior}.
\newblock \bibinfo{journal}{\emph{Proceedings of the National Academy of
  Sciences}} \bibinfo{volume}{118}, \bibinfo{number}{27}
  (\bibinfo{year}{2021}).
\newblock
\urldef\tempurl%
\url{https://doi.org/10.1073/pnas.2025764118}
\showDOI{\tempurl}


\bibitem[Baum(2020)]%
        {BaumMidTerm}
\bibfield{author}{\bibinfo{person}{Seth~D. Baum}.}
  \bibinfo{year}{2020}\natexlab{}.
\newblock \showarticletitle{Medium-Term Artificial Intelligence and Society}.
\newblock \bibinfo{journal}{\emph{Information}} \bibinfo{volume}{11},
  \bibinfo{number}{6} (\bibinfo{year}{2020}).
\newblock
\urldef\tempurl%
\url{https://doi.org/10.3390/info11060290}
\showDOI{\tempurl}


\bibitem[Baum et~al\mbox{.}(2019)]%
        {LongTermTraj}
\bibfield{author}{\bibinfo{person}{Seth~D. Baum}, \bibinfo{person}{Stuart
  Armstrong}, \bibinfo{person}{Timoteus Ekenstedt}, \bibinfo{person}{Olle
  H\"aggstr\"om}, \bibinfo{person}{Robin Hanson}, \bibinfo{person}{Karin
  Kuhlemann}, \bibinfo{person}{Matthijs~M. Maas}, \bibinfo{person}{James~D.
  Miller}, \bibinfo{person}{Markus Salmela}, \bibinfo{person}{Anders Sandberg},
  \bibinfo{person}{Kaj Sotala}, \bibinfo{person}{Phil Torres},
  \bibinfo{person}{Alexey Turchin}, {and} \bibinfo{person}{Roman~V.
  Yampolskiy}.} \bibinfo{year}{2019}\natexlab{}.
\newblock \showarticletitle{Long-term trajectories of human civilization}.
\newblock \bibinfo{journal}{\emph{Foresight}} \bibinfo{volume}{21},
  \bibinfo{number}{1} (\bibinfo{year}{2019}), \bibinfo{pages}{53--83}.
\newblock
\urldef\tempurl%
\url{https://doi.org/10.1108/FS-04-2018-0037}
\showDOI{\tempurl}


\bibitem[{BBC News}(2021)]%
        {AusVGoogle}
\bibfield{author}{\bibinfo{person}{{BBC News}}.}
  \bibinfo{year}{2021}\natexlab{}.
\newblock \bibinfo{title}{Australia News Code: {W}hat's this row with
  {F}acebook and {G}oogle all about?}
\newblock
  \bibinfo{howpublished}{\url{https://www.bbc.com/news/world-australia-56107028}}.
\newblock
\newblock
\shownote{Accessed: 2021-08-26}.


\bibitem[Belkhir and Elmeligi(2018)]%
        {AIEmissions}
\bibfield{author}{\bibinfo{person}{Lotfi Belkhir} {and} \bibinfo{person}{Ahmed
  Elmeligi}.} \bibinfo{year}{2018}\natexlab{}.
\newblock \showarticletitle{Assessing {ICT} global emissions footprint:
  {T}rends to 2040 \& recommendations}.
\newblock \bibinfo{journal}{\emph{Journal of Cleaner Production}}
  \bibinfo{volume}{177} (\bibinfo{year}{2018}), \bibinfo{pages}{448--463}.
\newblock
\urldef\tempurl%
\url{https://doi.org/10.1016/j.jclepro.2017.12.239}
\showDOI{\tempurl}


\bibitem[Born et~al\mbox{.}(1955)]%
        {RussellEinstein}
\bibfield{author}{\bibinfo{person}{Max Born}, \bibinfo{person}{Percy~W.
  Bridgman}, \bibinfo{person}{Albert Einstein}, \bibinfo{person}{Leopold
  Infeld}, \bibinfo{person}{Frederic Joliot-Curie}, \bibinfo{person}{Herman~J.
  Muller}, \bibinfo{person}{Linus Pauling}, \bibinfo{person}{Cecil~F. Powell},
  \bibinfo{person}{Joseph Rotblat}, \bibinfo{person}{Bertrand Russell}, {and}
  \bibinfo{person}{Hideki Yukawa}.} \bibinfo{year}{1955}\natexlab{}.
\newblock \bibinfo{title}{Russell-{E}instein Manifesto}.
\newblock \bibinfo{howpublished}{Available at
  \url{https://www.atomicheritage.org/key-documents/russell-einstein-manifesto}}.
\newblock
\newblock
\shownote{Accessed: 2022-03-01}.


\bibitem[Bostrom(2002)]%
        {BostromXRisk}
\bibfield{author}{\bibinfo{person}{Nick Bostrom}.}
  \bibinfo{year}{2002}\natexlab{}.
\newblock \showarticletitle{Existential Risks: Analyzing Human Extinction
  Scenarios and Related Hazards}.
\newblock \bibinfo{journal}{\emph{Journal of Evolution and Technology}}
  \bibinfo{volume}{9} (\bibinfo{year}{2002}).
\newblock
\urldef\tempurl%
\url{https://www.nickbostrom.com/existential/risks.pdf}
\showURL{%
\tempurl}


\bibitem[Bostrom(2014)]%
        {Superintelligence}
\bibfield{author}{\bibinfo{person}{Nick Bostrom}.}
  \bibinfo{year}{2014}\natexlab{}.
\newblock \bibinfo{booktitle}{\emph{{Superintelligence: Paths, Dangers,
  Strategies}}}.
\newblock \bibinfo{publisher}{Oxford University Press}.
\newblock
\showISBNx{978-0199678112}


\bibitem[Bostrom and \'Cirkovi\'c(2008)]%
        {Bostrom08GCRs}
\bibfield{author}{\bibinfo{person}{Nick Bostrom} {and}
  \bibinfo{person}{Milan~M. \'Cirkovi\'c}.} \bibinfo{year}{2008}\natexlab{}.
\newblock \bibinfo{booktitle}{\emph{Global Catastrophic Risks}}.
\newblock \bibinfo{publisher}{Oxford University Press}.
\newblock
\showISBNx{978-0199606504}


\bibitem[Boulanin et~al\mbox{.}(2019)]%
        {SIPRI19Vol1}
\bibfield{author}{\bibinfo{person}{Vincent Boulanin}, \bibinfo{person}{Shahar
  Avin}, \bibinfo{person}{Frank Sauer}, \bibinfo{person}{John Borrie},
  \bibinfo{person}{Dimitri Scheftelowitsch}, \bibinfo{person}{Justin Bronk},
  \bibinfo{person}{Page~O. Stoutland}, \bibinfo{person}{Martin Hagstr\"om},
  \bibinfo{person}{Petr Topychkanov}, \bibinfo{person}{Michael~C. Horowitz},
  \bibinfo{person}{Anja Kaspersen}, \bibinfo{person}{Chris King},
  \bibinfo{person}{S.M. Amadae}, {and} \bibinfo{person}{Jean-Marc Rickli}.}
  \bibinfo{year}{2019}\natexlab{}.
\newblock \bibinfo{booktitle}{\emph{The Impact of Artificial Intelligence on
  Strategic Stability and Nuclear Risk, Volume I, Euro-Atlantic Perspectives}}.
\newblock \bibinfo{type}{{T}echnical {R}eport}. \bibinfo{institution}{SIPRI}.
\newblock


\bibitem[Brundage et~al\mbox{.}(2018)]%
        {MaliciousAIReport}
\bibfield{author}{\bibinfo{person}{Miles Brundage}, \bibinfo{person}{Shahar
  Avin}, \bibinfo{person}{Jack Clark}, \bibinfo{person}{Helen Toner},
  \bibinfo{person}{Peter Eckersley}, \bibinfo{person}{Ben Garfinkel},
  \bibinfo{person}{Allan Dafoe}, \bibinfo{person}{Paul Scharre},
  \bibinfo{person}{Thomas Zeitzoff}, \bibinfo{person}{Bobby Filar},
  \bibinfo{person}{Hyrum Anderson}, \bibinfo{person}{Heather Roff},
  \bibinfo{person}{Gregory~C. Allen}, \bibinfo{person}{Jacob Steinhardt},
  \bibinfo{person}{Carrick Flynn}, \bibinfo{person}{Se\'{a}n
  \'{O}h\'{E}igeartaigh}, \bibinfo{person}{Simon Beard}, \bibinfo{person}{Haydn
  Belfield}, \bibinfo{person}{Sebastian Farquhar}, \bibinfo{person}{Clare
  Lyle}, \bibinfo{person}{Rebecca Crootof}, \bibinfo{person}{Owain Evans},
  \bibinfo{person}{Michael Page}, \bibinfo{person}{Joanna Bryson},
  \bibinfo{person}{Roman Yampolskiy}, {and} \bibinfo{person}{Dario Amodei}.}
  \bibinfo{year}{2018}\natexlab{}.
\newblock \bibinfo{booktitle}{\emph{The Malicious Use of Artificial
  Intelligence: Forecasting, Prevention, and Mitigation}}.
\newblock \bibinfo{type}{{T}echnical {R}eport}.
\newblock
\urldef\tempurl%
\url{https://maliciousaireport.com/}
\showURL{%
\tempurl}


\bibitem[Buchanan and Imbrie(2022)]%
        {Buchanan22NewFire}
\bibfield{author}{\bibinfo{person}{Ben Buchanan} {and} \bibinfo{person}{Andrew
  Imbrie}.} \bibinfo{year}{2022}\natexlab{}.
\newblock \bibinfo{booktitle}{\emph{The New Fire: War, Peace, and Democracy in
  the Age of AI}}.
\newblock \bibinfo{publisher}{MIT Press}.
\newblock
\showISBNx{9780262046541}


\bibitem[Calders et~al\mbox{.}(2021)]%
        {calders2021introduction}
\bibfield{author}{\bibinfo{person}{Toon Calders}, \bibinfo{person}{Eirini
  Ntoutsi}, \bibinfo{person}{Mykola Pechenizkiy}, \bibinfo{person}{Bodo
  Rosenhahn}, {and} \bibinfo{person}{Salvatore Ruggieri}.}
  \bibinfo{year}{2021}\natexlab{}.
\newblock \showarticletitle{Introduction to The Special Section on Bias and
  Fairness in AI}.
\newblock \bibinfo{journal}{\emph{ACM SIGKDD Explorations Newsletter}}
  \bibinfo{volume}{23}, \bibinfo{number}{1} (\bibinfo{year}{2021}),
  \bibinfo{pages}{1--3}.
\newblock


\bibitem[Callaway(2021)]%
        {ProteinFold}
\bibfield{author}{\bibinfo{person}{Ewen Callaway}.}
  \bibinfo{year}{2021}\natexlab{}.
\newblock \showarticletitle{{DeepMind}'s {AI} predicts structures for a vast
  trove of proteins}.
\newblock \bibinfo{journal}{\emph{{Nature}}}  \bibinfo{volume}{595}
  (\bibinfo{date}{Jul} \bibinfo{year}{2021}), \bibinfo{pages}{635}.
\newblock
\urldef\tempurl%
\url{https://doi.org/10.1038/d41586-021-02025-4}
\showDOI{\tempurl}


\bibitem[Carlsmith(2021)]%
        {Carlsmith}
\bibfield{author}{\bibinfo{person}{Joseph Carlsmith}.}
  \bibinfo{year}{2021}\natexlab{}.
\newblock \bibinfo{booktitle}{\emph{{Is power-seeking AI an existential
  risk?}}}
\newblock \bibinfo{type}{{T}echnical {R}eport}. \bibinfo{institution}{Open
  Philanthropy}.
\newblock
\newblock
\shownote{Draft report available at
  \url{https://www.lesswrong.com/posts/HduCjmXTBD4xYTegv/draft-report-on-existential-risk-from-power-seeking-ai}}.


\bibitem[Chesney and Citron(2019)]%
        {chesney2019deep}
\bibfield{author}{\bibinfo{person}{Bobby Chesney} {and}
  \bibinfo{person}{Danielle Citron}.} \bibinfo{year}{2019}\natexlab{}.
\newblock \showarticletitle{Deep fakes: A looming challenge for privacy,
  democracy, and national security}.
\newblock \bibinfo{journal}{\emph{Calif. L. Rev.}}  \bibinfo{volume}{107}
  (\bibinfo{year}{2019}), \bibinfo{pages}{1753}.
\newblock


\bibitem[Christian(2021)]%
        {Christian21Alignment}
\bibfield{author}{\bibinfo{person}{Brian Christian}.}
  \bibinfo{year}{2021}\natexlab{}.
\newblock \bibinfo{booktitle}{\emph{The Alignment Problem: How Can Machines
  Learn Human Values?}}
\newblock \bibinfo{publisher}{Atlantic Books Ltd.}
\newblock
\showISBNx{978-1-78649-430-6}


\bibitem[Christiano(2019)]%
        {WhatFailureLooksLike}
\bibfield{author}{\bibinfo{person}{Paul Christiano}.}
  \bibinfo{year}{2019}\natexlab{}.
\newblock \bibinfo{title}{What Failure Looks Like}.
\newblock
  \bibinfo{howpublished}{\url{https://www.lesswrong.com/posts/HBxe6wdjxK239zajf/what-failure-looks-like}}.
\newblock
\newblock
\shownote{Accessed: 2022-03-02}.


\bibitem[Chun(2021)]%
        {chun2021discriminating}
\bibfield{author}{\bibinfo{person}{Wendy Hui~Kyong Chun}.}
  \bibinfo{year}{2021}\natexlab{}.
\newblock \bibinfo{booktitle}{\emph{Discriminating Data: Correlation,
  Neighborhoods, and the New Politics of Recognition}}.
\newblock \bibinfo{publisher}{MIT Press}.
\newblock


\bibitem[Cinelli et~al\mbox{.}(2021)]%
        {EchoChamber}
\bibfield{author}{\bibinfo{person}{Matteo Cinelli}, \bibinfo{person}{Gianmarco
  De~Francisci~Morales}, \bibinfo{person}{Alessandro Galeazzi},
  \bibinfo{person}{Walter Quattrociocchi}, {and} \bibinfo{person}{Michele
  Starnini}.} \bibinfo{year}{2021}\natexlab{}.
\newblock \showarticletitle{The echo chamber effect on social media}.
\newblock \bibinfo{journal}{\emph{Proceedings of the National Academy of
  Sciences}} \bibinfo{volume}{118}, \bibinfo{number}{9} (\bibinfo{year}{2021}).
\newblock
\urldef\tempurl%
\url{https://doi.org/10.1073/pnas.2023301118}
\showDOI{\tempurl}


\bibitem[Cotton-Barratt et~al\mbox{.}(2020)]%
        {Cotton-Barratt20Defence}
\bibfield{author}{\bibinfo{person}{Owen Cotton-Barratt}, \bibinfo{person}{Max
  Daniel}, {and} \bibinfo{person}{Anders Sandberg}.}
  \bibinfo{year}{2020}\natexlab{}.
\newblock \showarticletitle{Defence in Depth Against Human Extinction:
  Prevention, Response, Resilience, and Why They All Matter}.
\newblock \bibinfo{journal}{\emph{Global Policy}} \bibinfo{volume}{11},
  \bibinfo{number}{3} (\bibinfo{year}{2020}), \bibinfo{pages}{271--282}.
\newblock
\urldef\tempurl%
\url{https://doi.org/10.1111/1758-5899.12786}
\showDOI{\tempurl}


\bibitem[Coupe et~al\mbox{.}(2019)]%
        {Coupe19NuclearWinter}
\bibfield{author}{\bibinfo{person}{Joshua Coupe}, \bibinfo{person}{Charles~G.
  Bardeen}, \bibinfo{person}{Alan Robock}, {and} \bibinfo{person}{Owen~B.
  Toon}.} \bibinfo{year}{2019}\natexlab{}.
\newblock \showarticletitle{Nuclear Winter Responses to Nuclear War Between the
  {United States} and {Russia} in the {Whole Atmosphere Community Climate Model
  Version 4} and the {Goddard Institute for Space Studies} {ModelE}}.
\newblock \bibinfo{journal}{\emph{Journal of Geophysical Research:
  Atmospheres}} \bibinfo{volume}{124}, \bibinfo{number}{15}
  (\bibinfo{year}{2019}), \bibinfo{pages}{8522--8543}.
\newblock
\urldef\tempurl%
\url{https://doi.org/10.1029/2019JD030509}
\showDOI{\tempurl}


\bibitem[Crawford(2021)]%
        {crawford2021atlas}
\bibfield{author}{\bibinfo{person}{Kate Crawford}.}
  \bibinfo{year}{2021}\natexlab{}.
\newblock \bibinfo{booktitle}{\emph{The Atlas of AI}}.
\newblock \bibinfo{publisher}{Yale University Press}.
\newblock


\bibitem[Crawford et~al\mbox{.}(2019)]%
        {AINOW2019}
\bibfield{author}{\bibinfo{person}{Kate Crawford}, \bibinfo{person}{Roel
  Dobbe}, \bibinfo{person}{Theodora Dryer}, \bibinfo{person}{Genevieve Fried},
  \bibinfo{person}{Ben Green}, \bibinfo{person}{Elizabeth Kaziunas},
  \bibinfo{person}{Amba Kak}, \bibinfo{person}{Varoon Mathur},
  \bibinfo{person}{Erin McElroy}, \bibinfo{person}{Andrea~Nill S\'anchez},
  \bibinfo{person}{Deborah Raji}, \bibinfo{person}{Joy~Lisi Rankin},
  \bibinfo{person}{Rashida Richardson}, \bibinfo{person}{Jason Schultz},
  \bibinfo{person}{Sarah~Myers West}, {and} \bibinfo{person}{Meredith
  Whittaker}.} \bibinfo{year}{2019}\natexlab{}.
\newblock \bibinfo{booktitle}{\emph{{AI Now 2019 Report}}}.
\newblock \bibinfo{type}{{T}echnical {R}eport}. \bibinfo{institution}{New York:
  AI Now Institute}.
\newblock
\urldef\tempurl%
\url{https://ainowinstitute.org/AI_Now_2019_Report.html}
\showURL{%
\tempurl}


\bibitem[Dafoe(2020)]%
        {DafoeAIGov}
\bibfield{author}{\bibinfo{person}{Allan Dafoe}.}
  \bibinfo{year}{2020}\natexlab{}.
\newblock \bibinfo{title}{{AI} Governance: {O}ppotunity and Theory of Impact}.
\newblock \bibinfo{howpublished}{\url{https://www.allandafoe.com/opportunity}}.
\newblock
\newblock
\shownote{Accessed: 2021-08-26}.


\bibitem[Dori-Hacohen et~al\mbox{.}(2021)]%
        {dorihacohen2021restoring}
\bibfield{author}{\bibinfo{person}{Shiri Dori-Hacohen}, \bibinfo{person}{Keen
  Sung}, \bibinfo{person}{Jengyu Chou}, {and} \bibinfo{person}{Julian
  Lustig-Gonzalez}.} \bibinfo{year}{2021}\natexlab{}.
\newblock \bibinfo{booktitle}{\emph{Restoring Healthy Online Discourse by
  Detecting and Reducing Controversy, Misinformation, and Toxicity Online}}.
\newblock \bibinfo{publisher}{Association for Computing Machinery},
  \bibinfo{address}{New York, NY, USA}, \bibinfo{pages}{2627--2628}.
\newblock
\showISBNx{9781450380379}


\bibitem[Drexler(2019)]%
        {DrexlerReframing}
\bibfield{author}{\bibinfo{person}{K.~Eric Drexler}.}
  \bibinfo{year}{2019}\natexlab{}.
\newblock \bibinfo{booktitle}{\emph{{Reframing Superintelligence: Comprehensive
  AI Services as General Intelligence}}}.
\newblock \bibinfo{type}{{T}echnical {R}eport}. \bibinfo{institution}{Future of
  Humanity Institute}.
\newblock


\bibitem[Duddu(2018)]%
        {duddu2018survey}
\bibfield{author}{\bibinfo{person}{Vasisht Duddu}.}
  \bibinfo{year}{2018}\natexlab{}.
\newblock \showarticletitle{A survey of adversarial machine learning in cyber
  warfare}.
\newblock \bibinfo{journal}{\emph{Defence Science Journal}}
  \bibinfo{volume}{68}, \bibinfo{number}{4} (\bibinfo{year}{2018}),
  \bibinfo{pages}{356}.
\newblock


\bibitem[Everitt et~al\mbox{.}(2018)]%
        {Everitt18AGISafety}
\bibfield{author}{\bibinfo{person}{Tom Everitt}, \bibinfo{person}{Gary Lea},
  {and} \bibinfo{person}{Marcus Hutter}.} \bibinfo{year}{2018}\natexlab{}.
\newblock \bibinfo{title}{{AGI} Safety Literature Review}.
\newblock
\newblock
\showeprint[arxiv]{1805.01109}~[cs.AI]


\bibitem[Favaro(2021)]%
        {Favaro21Distortion}
\bibfield{author}{\bibinfo{person}{Marina Favaro}.}
  \bibinfo{year}{2021}\natexlab{}.
\newblock \bibinfo{booktitle}{\emph{Weapons of Mass Distortion: A new approach
  to emerging technologies, risk reduction, and the global nuclear order}}.
\newblock \bibinfo{type}{{T}echnical {R}eport}. \bibinfo{institution}{Centre
  for Science and Security Studies}.
\newblock
\urldef\tempurl%
\url{https://www.kcl.ac.uk/csss/assets/weapons-of-mass-distortion.pdf}
\showURL{%
\tempurl}


\bibitem[Feldstein(2019)]%
        {FeldsteinSurveillance}
\bibfield{author}{\bibinfo{person}{Steven Feldstein}.}
  \bibinfo{year}{2019}\natexlab{}.
\newblock \bibinfo{title}{The Global Expansion of {AI} Surveillance}.
\newblock
  \bibinfo{howpublished}{\url{https://carnegieendowment.org/files/WP-Feldstein-AISurveillance\_final1.html}}.
\newblock
\newblock
\shownote{Accessed: 2021-08-26}.


\bibitem[for Humane~Technology(2021)]%
        {ledger2020}
\bibfield{author}{\bibinfo{person}{Center for Humane~Technology}.}
  \bibinfo{year}{2021}\natexlab{}.
\newblock \bibinfo{title}{Ledger of Harms}.
\newblock \bibinfo{howpublished}{\url{https://ledger.humanetech.com/}}.
\newblock


\bibitem[Garfinkel and Dafoe(2019)]%
        {Garfinkel19Scale}
\bibfield{author}{\bibinfo{person}{Ben Garfinkel} {and} \bibinfo{person}{Allan
  Dafoe}.} \bibinfo{year}{2019}\natexlab{}.
\newblock \showarticletitle{How does the offense-defense balance scale?}
\newblock \bibinfo{journal}{\emph{Journal of Strategic Studies}}
  \bibinfo{volume}{42}, \bibinfo{number}{6} (\bibinfo{year}{2019}),
  \bibinfo{pages}{736--763}.
\newblock
\urldef\tempurl%
\url{https://doi.org/10.1080/01402390.2019.1631810}
\showDOI{\tempurl}


\bibitem[Giger(2020)]%
        {Giger}
\bibfield{author}{\bibinfo{person}{Peter Giger}.}
  \bibinfo{year}{2020}\natexlab{}.
\newblock \bibinfo{title}{{COVID-19} could distract the world from even greater
  threats}.
\newblock
  \bibinfo{howpublished}{\url{https://www.weforum.org/agenda/2020/10/covid-19-distract-world-greater-threats/}}.
\newblock
\newblock
\shownote{Accessed: 2021-08-26}.


\bibitem[Gligori\'c et~al\mbox{.}(2019)]%
        {Brevity}
\bibfield{author}{\bibinfo{person}{Kristina Gligori\'c},
  \bibinfo{person}{Ashton Anderson}, {and} \bibinfo{person}{Robert West}.}
  \bibinfo{year}{2019}\natexlab{}.
\newblock \showarticletitle{Causal Effects of Brevity on Style and Success in
  Social Media}.
\newblock \bibinfo{journal}{\emph{Proc. ACM Hum.-Comput. Interact.}}
  \bibinfo{volume}{3}, \bibinfo{number}{CSCW}, Article \bibinfo{articleno}{45}
  (\bibinfo{year}{2019}), \bibinfo{numpages}{23}~pages.
\newblock
\urldef\tempurl%
\url{https://doi.org/10.1145/3359147}
\showDOI{\tempurl}


\bibitem[Gruetzemacher and Whittlestone(2022)]%
        {Gruetzemacher22Transformative}
\bibfield{author}{\bibinfo{person}{Ross Gruetzemacher} {and}
  \bibinfo{person}{Jess Whittlestone}.} \bibinfo{year}{2022}\natexlab{}.
\newblock \showarticletitle{The transformative potential of artificial
  intelligence}.
\newblock \bibinfo{journal}{\emph{Futures}}  \bibinfo{volume}{135}
  (\bibinfo{year}{2022}).
\newblock
\showISSN{0016-3287}
\urldef\tempurl%
\url{https://www.sciencedirect.com/science/article/pii/S0016328721001932}
\showURL{%
\tempurl}


\bibitem[Haskins(2019)]%
        {AmazonCoaching}
\bibfield{author}{\bibinfo{person}{Caroline Haskins}.}
  \bibinfo{year}{2019}\natexlab{}.
\newblock \bibinfo{title}{Amazon {I}s {C}oaching {C}ops on {H}ow to {O}btain
  {S}urveillance {F}ootage {W}ithout a {W}arrant}.
\newblock
  \bibinfo{howpublished}{\url{https://www.vice.com/en/article/43kga3/amazon-is-coaching-cops-on-how-to-obtain-surveillance-footage-without-a-warrant}}.
\newblock
\newblock
\shownote{Accessed: 2021-08-26}.


\bibitem[Hendrycks et~al\mbox{.}(2021)]%
        {Hendrycks21Unsolved}
\bibfield{author}{\bibinfo{person}{Dan Hendrycks}, \bibinfo{person}{Nicholas
  Carlini}, \bibinfo{person}{John Schulman}, {and} \bibinfo{person}{Jacob
  Steinhardt}.} \bibinfo{year}{2021}\natexlab{}.
\newblock \bibinfo{title}{Unsolved Problems in {ML} Safety}.
\newblock
\newblock
\showeprint[arxiv]{2109.13916}~[cs.LG]


\bibitem[Hwang(2020)]%
        {hwang2020subprime}
\bibfield{author}{\bibinfo{person}{Tim Hwang}.}
  \bibinfo{year}{2020}\natexlab{}.
\newblock \bibinfo{booktitle}{\emph{Subprime Attention Crisis: Advertising and
  the Time Bomb at the Heart of the Internet}}.
\newblock \bibinfo{publisher}{FSG originals}.
\newblock


\bibitem[Ingelsby and Adalja(2019)]%
        {Inglesby19GCBRs}
\bibfield{author}{\bibinfo{person}{Thomas~V. Ingelsby} {and}
  \bibinfo{person}{Amesh~A. Adalja}.} \bibinfo{year}{2019}\natexlab{}.
\newblock \bibinfo{booktitle}{\emph{Global Catastrophic Biological Risks}}.
\newblock \bibinfo{publisher}{Springer}.
\newblock
\showISBNx{978-3-030-36310-9}


\bibitem[Jaidka et~al\mbox{.}(2019)]%
        {TwitterDoubleLimit}
\bibfield{author}{\bibinfo{person}{Kokil Jaidka}, \bibinfo{person}{Alvin Zhou},
  {and} \bibinfo{person}{Yphtach Lelkes}.} \bibinfo{year}{2019}\natexlab{}.
\newblock \showarticletitle{Brevity is the Soul of Twitter: The Constraint
  Affordance and Political Discussion.}
\newblock \bibinfo{journal}{\emph{Journal of Communication}}
  \bibinfo{volume}{69}, \bibinfo{number}{4} (\bibinfo{year}{2019}),
  \bibinfo{pages}{345 -- 372}.
\newblock
\urldef\tempurl%
\url{https://doi.org/10.1093/joc/jqz023}
\showDOI{\tempurl}


\bibitem[Jhaver et~al\mbox{.}(2018)]%
        {jhaver2018online}
\bibfield{author}{\bibinfo{person}{Shagun Jhaver}, \bibinfo{person}{Sucheta
  Ghoshal}, \bibinfo{person}{Amy Bruckman}, {and} \bibinfo{person}{Eric
  Gilbert}.} \bibinfo{year}{2018}\natexlab{}.
\newblock \showarticletitle{Online Harassment and Content Moderation: The Case
  of Blocklists}.
\newblock \bibinfo{journal}{\emph{ACM Trans. Comput.-Hum. Interact.}}
  \bibinfo{volume}{25}, \bibinfo{number}{2} (\bibinfo{year}{2018}),
  \bibinfo{pages}{1--33}.
\newblock
\urldef\tempurl%
\url{https://doi.org/10.1145/3185593}
\showDOI{\tempurl}


\bibitem[Johnson(2019)]%
        {JohnsonAI}
\bibfield{author}{\bibinfo{person}{James Johnson}.}
  \bibinfo{year}{2019}\natexlab{}.
\newblock \showarticletitle{Artificial intelligence \& future warfare:
  implications for international security}.
\newblock \bibinfo{journal}{\emph{Defense \& Security Analysis}}
  \bibinfo{volume}{35}, \bibinfo{number}{2} (\bibinfo{year}{2019}),
  \bibinfo{pages}{147--169}.
\newblock
\urldef\tempurl%
\url{https://doi.org/10.1080/14751798.2019.1600800}
\showDOI{\tempurl}


\bibitem[Kaack et~al\mbox{.}(2021)]%
        {kaack2021aligning}
\bibfield{author}{\bibinfo{person}{Lynn Kaack}, \bibinfo{person}{Priya Donti},
  \bibinfo{person}{Emma Strubell}, \bibinfo{person}{George Kamiya},
  \bibinfo{person}{Felix Creutzig}, {and} \bibinfo{person}{David Rolnick}.}
  \bibinfo{year}{2021}\natexlab{}.
\newblock \showarticletitle{Aligning artificial intelligence with climate
  change mitigation}.
\newblock  (\bibinfo{year}{2021}).
\newblock
\newblock
\shownote{hal-03368037}.


\bibitem[Kaack et~al\mbox{.}(2020)]%
        {kaack2020artificial}
\bibfield{author}{\bibinfo{person}{Lynn Kaack}, \bibinfo{person}{Priya Donti},
  \bibinfo{person}{Emma Strubell}, {and} \bibinfo{person}{David Rolnick}.}
  \bibinfo{year}{2020}\natexlab{}.
\newblock \bibinfo{title}{Artificial Intelligence and Climate Change:
  Opportunities, considerations, and policy levers to align AI with climate
  change goals}.
\newblock
\newblock
\urldef\tempurl%
\url{https://eu.boell.org/en/2020/12/03/artificial-intelligence-and-climate-change}
\showURL{%
\tempurl}
\newblock
\shownote{Heinrich-B\"oll-Stiftung, Ecology}.


\bibitem[Kalimeris et~al\mbox{.}(2021)]%
        {kalimeris2021preference}
\bibfield{author}{\bibinfo{person}{Dimitris Kalimeris}, \bibinfo{person}{Smriti
  Bhagat}, \bibinfo{person}{Shankar Kalyanaraman}, {and} \bibinfo{person}{Udi
  Weinsberg}.} \bibinfo{year}{2021}\natexlab{}.
\newblock \showarticletitle{Preference Amplification in Recommender Systems}.
  In \bibinfo{booktitle}{\emph{Proceedings of the 27th ACM SIGKDD Conference on
  Knowledge Discovery \& Data Mining}} (Virtual Event, Singapore)
  \emph{(\bibinfo{series}{KDD '21})}. \bibinfo{publisher}{Association for
  Computing Machinery}, \bibinfo{address}{New York, NY, USA},
  \bibinfo{pages}{805--815}.
\newblock
\urldef\tempurl%
\url{https://doi.org/10.1145/3447548.3467298}
\showDOI{\tempurl}


\bibitem[Kaufman(2020)]%
        {BP}
\bibfield{author}{\bibinfo{person}{Mark Kaufman}.}
  \bibinfo{year}{2020}\natexlab{}.
\newblock \bibinfo{title}{The carbon footprint sham}.
\newblock
  \bibinfo{howpublished}{\url{https://mashable.com/feature/carbon-footprint-pr-campaign-sham}}.
\newblock
\newblock
\shownote{Accessed: 2021-08-26}.


\bibitem[Khanna(2019)]%
        {FutureAsian}
\bibfield{author}{\bibinfo{person}{Parag Khanna}.}
  \bibinfo{year}{2019}\natexlab{}.
\newblock \bibinfo{booktitle}{\emph{The {F}uture is {A}sian: {C}ommerce,
  {C}onflict, and {C}ulture in the 21st {C}entury}}.
\newblock \bibinfo{publisher}{{S}imon \& {S}chuster}.
\newblock


\bibitem[Krueger et~al\mbox{.}(2019)]%
        {Krueger2019}
\bibfield{author}{\bibinfo{person}{David Krueger}, \bibinfo{person}{Tegan
  Maharaj}, \bibinfo{person}{Shane Legg}, {and} \bibinfo{person}{Jan Leike}.}
  \bibinfo{year}{2019}\natexlab{}.
\newblock \showarticletitle{{Misleading Meta-Objectives and Hidden Incentives
  for Distributional Shift}}.
\newblock \bibinfo{journal}{\emph{Workshop on Safe Machine Learning at the 7th
  International Conference on Learning Representations (ICLR 2019)}}
  (\bibinfo{year}{2019}), \bibinfo{pages}{1--7}.
\newblock


\bibitem[Leslie(1996)]%
        {Leslie96End}
\bibfield{author}{\bibinfo{person}{John Leslie}.}
  \bibinfo{year}{1996}\natexlab{}.
\newblock \bibinfo{booktitle}{\emph{The End of the World: The Science and
  Ethics of Human Extinction}}.
\newblock \bibinfo{publisher}{Routledge}.
\newblock
\showISBNx{978-0415184472}


\bibitem[Lin(2019)]%
        {HerbertLin}
\bibfield{author}{\bibinfo{person}{Herbert Lin}.}
  \bibinfo{year}{2019}\natexlab{}.
\newblock \showarticletitle{The existential threat from cyber-enabled
  information warfare}.
\newblock \bibinfo{journal}{\emph{Bulletin of the Atomic Scientists}}
  \bibinfo{volume}{75}, \bibinfo{number}{4} (\bibinfo{year}{2019}),
  \bibinfo{pages}{187--196}.
\newblock
\urldef\tempurl%
\url{https://doi.org/10.1080/00963402.2019.1629574}
\showDOI{\tempurl}


\bibitem[Liu et~al\mbox{.}(2018)]%
        {liu2018governing}
\bibfield{author}{\bibinfo{person}{Hin-Yan Liu},
  \bibinfo{person}{Kristian~Cedervall Lauta}, {and}
  \bibinfo{person}{Matthijs~Michiel Maas}.} \bibinfo{year}{2018}\natexlab{}.
\newblock \showarticletitle{Governing Boring Apocalypses: A new typology of
  existential vulnerabilities and exposures for existential risk research}.
\newblock \bibinfo{journal}{\emph{Futures}}  \bibinfo{volume}{102}
  (\bibinfo{year}{2018}), \bibinfo{pages}{6--19}.
\newblock


\bibitem[Lotz(2019)]%
        {lotz2019amazon}
\bibfield{author}{\bibinfo{person}{Amanda Lotz}.}
  \bibinfo{year}{2019}\natexlab{}.
\newblock \bibinfo{title}{Amazon, Google and Facebook warrant antitrust
  scrutiny for many reasons -- not just because they're large}.
\newblock
\newblock
\urldef\tempurl%
\url{https://theconversation.com/amazon-google-and-facebook-warrant-antitrust-scrutiny-for-many-reasons-not-just-because-theyre-large-118370}
\showURL{%
\tempurl}


\bibitem[Lyons(2021)]%
        {Ring2000}
\bibfield{author}{\bibinfo{person}{Kim Lyons}.}
  \bibinfo{year}{2021}\natexlab{}.
\newblock \bibinfo{title}{Amazon's {R}ing now reportedly partners with more
  than 2,000 {US} police and fire departments}.
\newblock
  \bibinfo{howpublished}{\url{https://www.theverge.com/2021/1/31/22258856/amazon-ring-partners-police-fire-security-privacy-cameras}}.
\newblock
\newblock
\shownote{Accessed: 2021-08-26}.


\bibitem[Maas et~al\mbox{.}(2022)]%
        {Maas2022MilitaryAI}
\bibfield{author}{\bibinfo{person}{Matthijs~M. Maas}, \bibinfo{person}{Kayla
  Matteuci}, {and} \bibinfo{person}{Di Cooke}.}
  \bibinfo{year}{2022}\natexlab{}.
\newblock \showarticletitle{Military Artificial Intelligence as Contributor to
  Global Catastrophic Risk}. In \bibinfo{booktitle}{\emph{Cambridge Conference
  on Catastrophic Risks 2020}}.
\newblock
\newblock
\shownote{Draft available at
  \url{https://papers.ssrn.com/sol3/papers.cfm?abstract_id=4115010}}.


\bibitem[{Markets \& Markets}(2019)]%
        {AICyberMarket}
\bibfield{author}{\bibinfo{person}{{Markets \& Markets}}.}
  \bibinfo{year}{2019}\natexlab{}.
\newblock \bibinfo{title}{{AI} in Cybersecurity Market}.
\newblock
  \bibinfo{howpublished}{\url{https://www.marketsandmarkets.com/market-reports/ai-in-cybersecurity-market-224437074.html}}.
\newblock
\newblock
\shownote{Accessed: 2021-08-26}.


\bibitem[Morrison(2019)]%
        {ChinaGrowth}
\bibfield{author}{\bibinfo{person}{Wayne~M. Morrison}.}
  \bibinfo{year}{2019}\natexlab{}.
\newblock \bibinfo{booktitle}{\emph{China's {E}conomic {R}ise: {H}istory,
  Trends, Challenges, and Implications for the {U}nited {S}tates}}.
\newblock \bibinfo{type}{{T}echnical {R}eport}.
  \bibinfo{institution}{{C}ongressional {R}esearch {S}ervice}.
\newblock


\bibitem[Mueller(2019)]%
        {mueller2019mueller}
\bibfield{author}{\bibinfo{person}{Robert~S Mueller}.}
  \bibinfo{year}{2019}\natexlab{}.
\newblock \bibinfo{booktitle}{\emph{The Mueller report: Report on the
  investigation into Russian interference in the 2016 presidential election}}.
\newblock \bibinfo{publisher}{WSBLD}.
\newblock


\bibitem[{OECD}(2021)]%
        {OECD}
\bibfield{author}{\bibinfo{person}{{OECD}}.} \bibinfo{year}{2021}\natexlab{}.
\newblock \bibinfo{title}{130 countries and jurisdictions join bold new
  framework for international tax reform}.
\newblock
  \bibinfo{howpublished}{\url{https://www.oecd.org/newsroom/130-countries-and-jurisdictions-join-bold-new-framework-for-international-tax-reform.htm}}.
\newblock
\newblock
\shownote{Accessed: 2021-08-26}.


\bibitem[Ong and Caba{\~n}es(2018)]%
        {ong2018architects}
\bibfield{author}{\bibinfo{person}{Jonathan~Corpus Ong} {and}
  \bibinfo{person}{Jason Vincent~A. Caba{\~n}es}.}
  \bibinfo{year}{2018}\natexlab{}.
\newblock \bibinfo{booktitle}{\emph{Architects of networked disinformation:
  Behind the scenes of troll accounts and fake news production in the
  Philippines}}.
\newblock
\urldef\tempurl%
\url{https://doi.org/10.7275/2cq4-5396}
\showURL{%
\tempurl}


\bibitem[Ord(2020)]%
        {Precipice}
\bibfield{author}{\bibinfo{person}{Toby Ord}.} \bibinfo{year}{2020}\natexlab{}.
\newblock \bibinfo{booktitle}{\emph{{The Precipice: Existential Risk and the
  Future of Humanity}}}.
\newblock \bibinfo{publisher}{Hachette Books}.
\newblock
\showISBNx{978-0316484916}


\bibitem[Oreskes and Conway(2010)]%
        {oreskes2010merchants}
\bibfield{author}{\bibinfo{person}{Naomi Oreskes} {and}
  \bibinfo{person}{Erik~M. Conway}.} \bibinfo{year}{2010}\natexlab{}.
\newblock \bibinfo{booktitle}{\emph{Merchants of Doubt: How a Handful of
  Scientists Obscured the Truth on Issues from Tobacco Smoke to Global
  Warming}}.
\newblock \bibinfo{publisher}{Bloomsbury Publishing}.
\newblock
\showISBNx{9781608192939}


\bibitem[Pariser(2011)]%
        {pariser2011filter}
\bibfield{author}{\bibinfo{person}{Eli Pariser}.}
  \bibinfo{year}{2011}\natexlab{}.
\newblock \bibinfo{booktitle}{\emph{The Filter Bubble: How the New Personalized
  Web Is Changing What We Read and How We Think}}.
\newblock \bibinfo{publisher}{Penguin}.
\newblock


\bibitem[Posner(2004)]%
        {Posner04Catastrophe}
\bibfield{author}{\bibinfo{person}{Richard Posner}.}
  \bibinfo{year}{2004}\natexlab{}.
\newblock \bibinfo{booktitle}{\emph{Catastrophe: Risk and Response}}.
\newblock \bibinfo{publisher}{Oxford University Press}.
\newblock
\showISBNx{978-0195306477}


\bibitem[Prunkl and Whittlestone(2020)]%
        {Prunkl20Beyond}
\bibfield{author}{\bibinfo{person}{Carina Prunkl} {and} \bibinfo{person}{Jess
  Whittlestone}.} \bibinfo{year}{2020}\natexlab{}.
\newblock \showarticletitle{Beyond Near- and Long-Term: Towards a Clearer
  Account of Research Priorities in AI Ethics and Society}. In
  \bibinfo{booktitle}{\emph{Proceedings of the AAAI/ACM Conference on AI,
  Ethics, and Society}}. \bibinfo{publisher}{Association for Computing
  Machinery}, \bibinfo{address}{New York, NY, USA}, \bibinfo{pages}{138--143}.
\newblock
\urldef\tempurl%
\url{https://doi.org/10.1145/3375627.3375803}
\showURL{%
\tempurl}


\bibitem[Rees(2003)]%
        {Rees03OurFinalHour}
\bibfield{author}{\bibinfo{person}{Martin Rees}.}
  \bibinfo{year}{2003}\natexlab{}.
\newblock \bibinfo{booktitle}{\emph{Our Final Century: Will the Human Race
  Survive the Twenty-First Century?}}
\newblock \bibinfo{publisher}{William Heinemann}.
\newblock
\showISBNx{0-434-00809-5}


\bibitem[Rohith and Batth(2019)]%
        {rohith2019cyber}
\bibfield{author}{\bibinfo{person}{Cheerala Rohith} {and}
  \bibinfo{person}{Ranbir~Singh Batth}.} \bibinfo{year}{2019}\natexlab{}.
\newblock \showarticletitle{Cyber Warfare: Nations Cyber Conflicts, Cyber Cold
  War Between Nations and its Repercussion}. In \bibinfo{booktitle}{\emph{2019
  International Conference on Computational Intelligence and Knowledge Economy
  (ICCIKE)}}. IEEE, \bibinfo{pages}{640--645}.
\newblock


\bibitem[Rolnick et~al\mbox{.}(2022)]%
        {Rolnick22Tackling}
\bibfield{author}{\bibinfo{person}{David Rolnick}, \bibinfo{person}{Priya~L.
  Donti}, \bibinfo{person}{Lynn~H. Kaack}, \bibinfo{person}{Kelly Kochanski},
  \bibinfo{person}{Alexandre Lacoste}, \bibinfo{person}{Kris Sankaran},
  \bibinfo{person}{Andrew~Slavin Ross}, \bibinfo{person}{Nikola
  Milojevic-Dupont}, \bibinfo{person}{Natasha Jaques}, \bibinfo{person}{Anna
  Waldman-Brown}, \bibinfo{person}{Alexandra~Sasha Luccioni},
  \bibinfo{person}{Tegan Maharaj}, \bibinfo{person}{Evan~D. Sherwin},
  \bibinfo{person}{S.~Karthik Mukkavilli}, \bibinfo{person}{Konrad~P. Kording},
  \bibinfo{person}{Carla~P. Gomes}, \bibinfo{person}{Andrew~Y. Ng},
  \bibinfo{person}{Demis Hassabis}, \bibinfo{person}{John~C. Platt},
  \bibinfo{person}{Felix Creutzig}, \bibinfo{person}{Jennifer Chayes}, {and}
  \bibinfo{person}{Yoshua Bengio}.} \bibinfo{year}{2022}\natexlab{}.
\newblock \showarticletitle{Tackling Climate Change with Machine Learning}.
\newblock \bibinfo{journal}{\emph{ACM Comput. Surv.}} \bibinfo{volume}{55},
  \bibinfo{number}{2} (\bibinfo{year}{2022}), \bibinfo{numpages}{96}~pages.
\newblock
\urldef\tempurl%
\url{https://doi.org/10.1145/3485128}
\showURL{%
\tempurl}


\bibitem[Russell(2019a)]%
        {HumanCompatible}
\bibfield{author}{\bibinfo{person}{Stuart Russell}.}
  \bibinfo{year}{2019}\natexlab{a}.
\newblock \bibinfo{booktitle}{\emph{{Human Compatible: Artificial Intelligence
  and the Problem of Control}}}.
\newblock \bibinfo{publisher}{Viking Books}.
\newblock
\showISBNx{978-0-525-55861-3}


\bibitem[Russell(2019b)]%
        {Russell}
\bibfield{author}{\bibinfo{person}{Stuart~J. Russell}.}
  \bibinfo{year}{2019}\natexlab{b}.
\newblock \bibinfo{title}{Stuart {J}. {R}ussell on Filter Bubbles and the
  Future of {A}rtificial {I}ntelligence}.
\newblock
  \bibinfo{howpublished}{\url{https://www.youtube.com/watch?v=ZkV7anCPfaY\&ab\_channel=LongNowFoundation}}.
\newblock
\newblock
\shownote{Accessed: 2021-08-26}.


\bibitem[Saalman et~al\mbox{.}(2019)]%
        {SIPRI19Vol2}
\bibfield{author}{\bibinfo{person}{Lora Saalman}, \bibinfo{person}{Hwang
  Ji-Hwan}, \bibinfo{person}{Su Fei}, \bibinfo{person}{Jiang Tianjiao},
  \bibinfo{person}{Vasily Kashin}, \bibinfo{person}{Kim Ji-Sun},
  \bibinfo{person}{Vadim Kozyulin}, \bibinfo{person}{Arie Koichi},
  \bibinfo{person}{Li Xiang}, \bibinfo{person}{Cai Cuihong},
  \bibinfo{person}{Liu Yangyue}, \bibinfo{person}{Hwang Il-Soon}, {and}
  \bibinfo{person}{Nishida Michiru}.} \bibinfo{year}{2019}\natexlab{}.
\newblock \bibinfo{booktitle}{\emph{The Impact of Artificial Intelligence on
  Strategic Stability and Nuclear Risk, Volume II, East Asian Perspectives}}.
\newblock \bibinfo{type}{{T}echnical {R}eport}. \bibinfo{institution}{SIPRI}.
\newblock


\bibitem[Schiff et~al\mbox{.}(2021)]%
        {schiff21liar}
\bibfield{author}{\bibinfo{person}{Kaylyn~Jackson Schiff},
  \bibinfo{person}{Daniel~S. Schiff}, {and} \bibinfo{person}{Nat{\'a}lia~S
  Bueno}.} \bibinfo{year}{2021}\natexlab{}.
\newblock \showarticletitle{The Liar's Dividend: The Impact of Deepfakes and
  Fake News on Trust in Political Discourse}.
\newblock  (\bibinfo{year}{2021}).
\newblock
\newblock
\shownote{\url{https://doi.org/10.17605/OSF.IO/QPXR8}}.


\bibitem[Schmidt et~al\mbox{.}(2021)]%
        {NSCAI}
\bibfield{author}{\bibinfo{person}{Eric Schmidt}, \bibinfo{person}{Robert
  Work}, \bibinfo{person}{Safra Catz}, \bibinfo{person}{Eric Horvitz},
  \bibinfo{person}{Steve Chien}, \bibinfo{person}{Andrew Jassy},
  \bibinfo{person}{Clyburn Mignon}, \bibinfo{person}{Gilman Louie},
  \bibinfo{person}{Chris Darby}, \bibinfo{person}{Willian Mark},
  \bibinfo{person}{Kenneth Ford}, \bibinfo{person}{Jason Matheny},
  \bibinfo{person}{Jos\'e-Marie Griffiths}, \bibinfo{person}{Katharina
  McFarland}, {and} \bibinfo{person}{Andrew Moore}.}
  \bibinfo{year}{2021}\natexlab{}.
\newblock \bibinfo{booktitle}{\emph{Final Report}}.
\newblock \bibinfo{type}{{T}echnical {R}eport}. \bibinfo{institution}{{National
  Security Commission on Artificial Intelligence}}.
\newblock
\urldef\tempurl%
\url{https://www.nscai.gov/wp-content/uploads/2021/03/Full-Report-Digital-1.pdf}
\showURL{%
\tempurl}


\bibitem[Schneier(2018)]%
        {schneier2018click}
\bibfield{author}{\bibinfo{person}{Bruce Schneier}.}
  \bibinfo{year}{2018}\natexlab{}.
\newblock \bibinfo{booktitle}{\emph{Click Here to Kill Everybody: Security and
  Survival in a Hyper-connected World}}.
\newblock \bibinfo{publisher}{W. W. Norton \& Company}.
\newblock


\bibitem[Seger et~al\mbox{.}(2020)]%
        {SegerEpistemicSecurity}
\bibfield{author}{\bibinfo{person}{Elizabeth Seger}, \bibinfo{person}{Shahar
  Avin}, \bibinfo{person}{Gavin Pearson}, \bibinfo{person}{Mark Briers},
  \bibinfo{person}{Se\'an \'O~h\'Eigeartaigh}, {and} \bibinfo{person}{Helena
  Bacon}.} \bibinfo{year}{2020}\natexlab{}.
\newblock \bibinfo{booktitle}{\emph{Tackling threats to informed
  decision-making in democratic societies: {P}romoting epistemic security in a
  technologicall-advanced world}}.
\newblock \bibinfo{type}{{T}echnical {R}eport}. \bibinfo{institution}{The Alan
  Turing Institute, Defence and Security Programme}.
\newblock
\urldef\tempurl%
\url{https://www.cser.ac.uk/resources/epistemic-security/}
\showURL{%
\tempurl}


\bibitem[Sharma et~al\mbox{.}(2022)]%
        {sharma2022dark}
\bibfield{author}{\bibinfo{person}{Bhakti Sharma}, \bibinfo{person}{Susanna~S.
  Lee}, {and} \bibinfo{person}{Benjamin~K. Johnson}.}
  \bibinfo{year}{2022}\natexlab{}.
\newblock \showarticletitle{The {Dark} at the {End} of the {Tunnel}:
  Doomscrolling on {Social} {Media} {Newsfeeds}}.
\newblock \bibinfo{journal}{\emph{Technology, Mind, and Behavior}}
  \bibinfo{volume}{3}, \bibinfo{number}{1} (\bibinfo{year}{2022}).
\newblock
\newblock
\shownote{https://tmb.apaopen.org/pub/nn9uaqsz}.


\bibitem[Strubell et~al\mbox{.}(2019)]%
        {strubell2019energy}
\bibfield{author}{\bibinfo{person}{Emma Strubell}, \bibinfo{person}{Ananya
  Ganesh}, {and} \bibinfo{person}{Andrew McCallum}.}
  \bibinfo{year}{2019}\natexlab{}.
\newblock \bibinfo{title}{Energy and Policy Considerations for Deep Learning in
  {NLP}}.
\newblock
\newblock
\showeprint[arxiv]{1906.02243}~[cs.CL]


\bibitem[Sunstein(2018)]%
        {sunstein2018republic}
\bibfield{author}{\bibinfo{person}{Cass~R. Sunstein}.}
  \bibinfo{year}{2018}\natexlab{}.
\newblock \bibinfo{booktitle}{\emph{\# Republic: Divided Democracy in the Age
  of Social Media}}.
\newblock \bibinfo{publisher}{Princeton University Press}.
\newblock


\bibitem[Taddeo et~al\mbox{.}(2019)]%
        {AICyberSword}
\bibfield{author}{\bibinfo{person}{Mariarosaria Taddeo}, \bibinfo{person}{Tom
  McCutcheon}, {and} \bibinfo{person}{Luciano Floridi}.}
  \bibinfo{year}{2019}\natexlab{}.
\newblock \showarticletitle{Trusting Artificial Intelligence in Cybersecurity
  is a Double-Edged Sword}.
\newblock \bibinfo{journal}{\emph{Nat Mach Intell}}  \bibinfo{volume}{1}
  (\bibinfo{year}{2019}), \bibinfo{pages}{557--560}.
\newblock
\urldef\tempurl%
\url{https://doi.org/10.1038/s42256-019-0109-1}
\showDOI{\tempurl}


\bibitem[Topychkanov et~al\mbox{.}(2030)]%
        {SIPRI20Vol3}
\bibfield{author}{\bibinfo{person}{Petr Topychkanov}, \bibinfo{person}{Kritika
  Roy}, \bibinfo{person}{Saima~Aman Sial}, \bibinfo{person}{Dmitry
  Stefanovich}, \bibinfo{person}{Maaike Verbruggen}, \bibinfo{person}{Sanatan
  Kulshrestha}, \bibinfo{person}{Yanitra Kumaraguru}, {and}
  \bibinfo{person}{Malinda Meegoda}.} \bibinfo{year}{2030}\natexlab{}.
\newblock \bibinfo{booktitle}{\emph{The Impact of Artificial Intelligence on
  Strategic Stability and Nuclear Risk, Volume III, South Asian Perspectives}}.
\newblock \bibinfo{type}{{T}echnical {R}eport}. \bibinfo{institution}{SIPRI}.
\newblock


\bibitem[Urbina et~al\mbox{.}(2022)]%
        {Urbina22DualUse}
\bibfield{author}{\bibinfo{person}{Fabio Urbina}, \bibinfo{person}{Filippa
  Lentzos}, \bibinfo{person}{C\'edric Invernizzi}, {and} \bibinfo{person}{Sean
  Ekins}.} \bibinfo{year}{2022}\natexlab{}.
\newblock \showarticletitle{Dual use of artificial-intelligence-powered drug
  discovery}.
\newblock \bibinfo{journal}{\emph{Nature Machine Intelligence}}
  \bibinfo{volume}{4} (\bibinfo{year}{2022}), \bibinfo{pages}{189--191}.
\newblock
\urldef\tempurl%
\url{https://doi.org/10.1038/s42256-022-00465-9}
\showURL{%
\tempurl}


\bibitem[Vicario et~al\mbox{.}(2016)]%
        {Vicario2016}
\bibfield{author}{\bibinfo{person}{Michela~Del Vicario},
  \bibinfo{person}{Alessandro Bessi}, \bibinfo{person}{Fabiana Zollo},
  \bibinfo{person}{Fabio Petroni}, \bibinfo{person}{Antonio Scala},
  \bibinfo{person}{Guido Caldarelli}, \bibinfo{person}{H.~Eugene Stanley},
  {and} \bibinfo{person}{Walter Quattrociocchi}.}
  \bibinfo{year}{2016}\natexlab{}.
\newblock \showarticletitle{The spreading of misinformation online}.
\newblock \bibinfo{journal}{\emph{Proceedings of the National Academy of
  Sciences}} \bibinfo{volume}{113}, \bibinfo{number}{3} (\bibinfo{year}{2016}),
  \bibinfo{pages}{554--559}.
\newblock
\urldef\tempurl%
\url{https://doi.org/10.1073/pnas.1517441113}
\showDOI{\tempurl}


\bibitem[Webster et~al\mbox{.}(2017)]%
        {AIDP}
\bibfield{author}{\bibinfo{person}{Graham Webster}, \bibinfo{person}{Rogier
  Creemers}, \bibinfo{person}{Paul Triolo}, {and} \bibinfo{person}{Elsa
  Kania}.} \bibinfo{year}{2017}\natexlab{}.
\newblock \bibinfo{title}{Full {T}ranslation: {C}hina's `{N}ew {G}eneration
  {A}rtificial {I}ntelligence {D}evelopment {P}lan' (2017)}.
\newblock
  \bibinfo{howpublished}{\url{https://www.newamerica.org/cybersecurity-initiative/digichina/blog/full-translation-chinas-new-generation-artificial-intelligence-development-plan-2017/}}.
\newblock
\newblock
\shownote{Accessed: 2021-08-26}.


\bibitem[Whittaker et~al\mbox{.}(2018)]%
        {AINOW2018}
\bibfield{author}{\bibinfo{person}{Meredith Whittaker}, \bibinfo{person}{Kate
  Crawford}, \bibinfo{person}{Roel Dobbe}, \bibinfo{person}{Genevieve Fried},
  \bibinfo{person}{Elizabeth Kaziunas}, \bibinfo{person}{Varoon Mathur},
  \bibinfo{person}{Sarah~Myers West}, \bibinfo{person}{Rashida Richardson},
  \bibinfo{person}{Jason Schultz}, {and} \bibinfo{person}{Oscar Schwartz}.}
  \bibinfo{year}{2018}\natexlab{}.
\newblock \bibinfo{booktitle}{\emph{{AI Now 2018 Report}}}.
\newblock \bibinfo{type}{{T}echnical {R}eport}. \bibinfo{institution}{New York:
  AI Now Institute}.
\newblock
\urldef\tempurl%
\url{https://ainowinstitute.org/AI_Now_2018_Report.html}
\showURL{%
\tempurl}


\bibitem[Whittlestone et~al\mbox{.}(2021)]%
        {Whittlestone21Societal}
\bibfield{author}{\bibinfo{person}{Jess Whittlestone}, \bibinfo{person}{Kai
  Arulkumaran}, {and} \bibinfo{person}{Matthew Crosby}.}
  \bibinfo{year}{2021}\natexlab{}.
\newblock \showarticletitle{The Societal Implications of Deep Reinforcement
  Learning}.
\newblock \bibinfo{journal}{\emph{J. Artif. Int. Res.}}  \bibinfo{volume}{70}
  (\bibinfo{year}{2021}), \bibinfo{pages}{1003--1030}.
\newblock
\showISSN{1076-9757}
\urldef\tempurl%
\url{https://doi.org/10.1613/jair.1.12360}
\showURL{%
\tempurl}


\bibitem[Wirkuttis and Klein(2017)]%
        {AIinCyberSec}
\bibfield{author}{\bibinfo{person}{Nadine Wirkuttis} {and}
  \bibinfo{person}{Hadas Klein}.} \bibinfo{year}{2017}\natexlab{}.
\newblock \showarticletitle{Artificial intelligence in cybersecurity}.
\newblock \bibinfo{journal}{\emph{Cyber, Intelligence, and Security}}
  \bibinfo{volume}{1}, \bibinfo{number}{1} (\bibinfo{year}{2017}),
  \bibinfo{pages}{103--119}.
\newblock


\bibitem[Zuboff(2019)]%
        {zuboff2019age}
\bibfield{author}{\bibinfo{person}{Shoshana Zuboff}.}
  \bibinfo{year}{2019}\natexlab{}.
\newblock \bibinfo{booktitle}{\emph{The Age of Surveillance Capitalism: The
  Fight for a Human Future at the New Frontier of Power}}.
\newblock \bibinfo{publisher}{Hachette UK}.
\newblock


\bibitem[Zwetsloot and Dafoe(2019)]%
        {Zwetsloot19Thinking}
\bibfield{author}{\bibinfo{person}{Remco Zwetsloot} {and}
  \bibinfo{person}{Allan Dafoe}.} \bibinfo{year}{2019}\natexlab{}.
\newblock \showarticletitle{Thinking About Risks From {AI}: Accidents, Misuse
  and Structure}.
\newblock \bibinfo{journal}{\emph{Lawfare}} (\bibinfo{year}{2019}).
\newblock
\newblock
\shownote{\url{https://www.lawfareblog.com/thinking-about-risks-ai-accidents-misuse-and-structure}}.


\end{thebibliography}


\end{document}